\documentclass{aa}  
\usepackage{graphicx}
\usepackage{float}
\usepackage{wrapfig}
\usepackage{ragged2e}
\usepackage{geometry}
\usepackage{txfonts}
\usepackage{dblfloatfix}
\usepackage{subcaption}
\usepackage{hyperref}
\usepackage{ulem}
\usepackage[dvipsnames]{xcolor}
\hypersetup{
    colorlinks = true,
    citecolor ={cyan}
}

\newcommand{\Msun}{\ensuremath{\rm \,M_\odot}}
\newcommand{\Msunyr}{\ensuremath{\rm \,M_\odot \, yr^{-1}}}

\newcommand{\Rsun}{\ensuremath{\rm \,R_\odot}}
\newcommand{\Zsun}{\ensuremath{\,Z_\odot}}

\bibpunct[; ]{(}{)}{;}{a}{}{;}

\definecolor{darkgreen}{rgb}{0.0, 0.4, 0.22}

\definecolor{ochre}{rgb}{0.8, 0.47, 0.13}

\newcommand{\Mpy}{\ensuremath{\,\rm M_{\odot}\, yr^{-1}}}

\newcommand{\yr}{\ensuremath{\,\rm yr}}

\usepackage{array}
\newcolumntype{L}[1]{>{\raggedright\let\newline\\\arraybackslash\hspace{0pt}}m{#1}}
\newcolumntype{C}[1]{>{\centering\let\newline\\\arraybackslash\hspace{0pt}}m{#1}}
\newcolumntype{R}[1]{>{\raggedleft\let\newline\\\arraybackslash\hspace{0pt}}m{#1}}

\begin{document} 
   \title{A fundamental limit to how close binary systems can get via stable mass transfer shapes the properties of binary black hole mergers}
   \titlerunning{Comfort zones of stars: a fundamental limit on orbital tightening via stable mass transfer}
   \authorrunning{Klencki et al.}
   
      \author{Jakub Klencki\inst{1,2}
          \and Philipp Podsiadlowski\inst{3}
          \and Norbert Langer\inst{4}
          \and Aleksandra Olejak\inst{1}
          \and Stephen Justham\inst{1}
          \and Alejandro Vigna-G\'omez\inst{1}
          \and Selma E. de Mink\inst{1}
   }

   \institute{
    Max Planck Institute for Astrophysics, Karl-Schwarzschild-Strasse 1, 85748 Garching, Germany\\
     \email{jklencki@mpa-garching.mpg.de}
    \and
   European Southern Observatory, Karl-Schwarzschild-Strasse 2, 85748 Garching bei München, Germany
     \and
   University of Oxford, St Edmund Hall, Oxford OX1 4AR, UK
   \and
   Argelander Institut für Astronomie,
              Auf dem Hügel 71, 53121 Bonn, Germany
   }
   \date{Received May, 2025; accepted...}
   
  \abstract
   {
Mass transfer in binary systems is the key process in the formation of various classes of objects, including merging binary black holes (BBHs) and neutron stars. Orbital evolution during mass transfer depends on how much mass is accreted and how much angular momentum is lost -- two of the main uncertainties in binary evolution. This poses a challenge for obtaining reliable predictions from binary channels.
Here, we demonstrate that, despite these unknowns, a fundamental limit exists to how close binary systems can get via stable mass transfer (SMT) that is robust against uncertainties in orbital evolution. Based on detailed evolutionary models of interacting systems with a BH accretor and a massive-star companion, we show that the post-interaction orbit is always wider than $\sim$10$\Rsun$, even when extreme shrinkage due to L2 outflows is assumed. Systems evolving towards tighter orbits become dynamically unstable and result in stellar mergers.
This separation limit has direct implications for the properties of BBH mergers: long delay times ($\gtrsim$1 Gyr), and an absence of high BH spins from the tidal spin-up of helium stars. At high metallicity, the SMT channel may be severely quenched due to Wolf-Rayet winds. We predict BBH mergers from $\sim$10$\Msun$ to 90$\Msun$, with case A mass transfer dominating above 40$\Msun$. 
The reason for the separation limit lies in the stellar structure, not in binary physics. If the orbit gets too narrow during mass transfer, a dynamical instability is triggered by a rapid expansion of the remaining donor envelope due to its near-flat entropy profile. The closest separations can be achieved from core-He burning ($\sim$8$-$15$\Rsun$) and Main Sequence donors ($\sim$15$-$30$\Rsun$), while Hertzsprung Gap donors lead to wider orbits ($\gtrsim$30$-$50$\Rsun$) and non-merging BBHs. 
These outcomes and mass transfer stability are determined by the entropy structures, which are governed by internal composition profiles. Consequently, the formation of BBH mergers and other compact binaries via SMT is a sensitive probe of chemical mixing in stars, and may help address open questions of stellar astrophysics, such as the blue-supergiant problem. 
Finally, we propose a new, simplified treatment of mass transfer stability that more accurately reproduces detailed results and remains flexible under varying assumptions for orbital evolution. 
   }
  
   \maketitle

\section{Introduction}
\label{sec.intro}

We are at the brink of the data-driven era of gravitational-wave (GW) astrophysics. With the three observing runs concluded and the forth run O4 currently underway, the LIGO-Virgo-Kagra collaboration has reported detections of about 100 compact-binary coalescences, most of which being binary black hole (BBH) mergers \citep{gwtc2_2020,gwtc3_LVK}. In the next few years, the number of GW detections will increase to thousands and eventually many millions with the increasing sensitivity of the current detectors \citep[O4-O5, AdV+][]{Abbott2020LRR} and the next-generation instruments such as Einstein Telescope \citep{Maggiore2020} and Cosmic Explorer \citep{Reitze2019}. With millions of compact binary coalescences detected every year, merging pairs of black holes (BH) and neutron stars (NS) will arguably become the best measured state in the evolution of massive stars across the Universe. 

GW astronomy has enormous potential for probing massive-star formation and evolution across all environments and redshifts. The key to unlocking this treasure chest are population models exploring formation scenarios for GW sources that allow for a confrontation between theory and observation. However, this undertaking is currently hindered by the Interpretation Challenge: due to severe uncertainties and multiple free parameters in the formation scenarios, we are unable to robustly interpret the observed population and link the GW sources with their progenitor stars \citep{Broekgaarden2022,MandelBroekgaarden2022,Chruslinska2024}. The task is not made easier by the fact that multiple diverse formation channels may simultaneously be at play \citep{Zevin2021}, such as isolated evolution of stellar multiples \citep{Belczynski2016,Tauris2017,Stevenson2017,Antonini2017,Kruckow2018,Mapelli2018,Belczynski2020,Vynatheya2022}, dynamical evolution in dense stellar clusters \citep{Rodriguez2016a,Askar2017,Samsing2018,Rodriguez2019,Mapelli2021,Mapelli2022}, evolution in gas-rich environments such as AGN disks \citep{Antonini2016,Stone2017,McKernan2018,Samsing2022}, or channels involving the existence of primordial BHs, formed shortly after the Big Bang \citep{Sasaki2018,Papanikolaou2021}. 

\begin{figure}[h!]
    \includegraphics[width=\columnwidth]{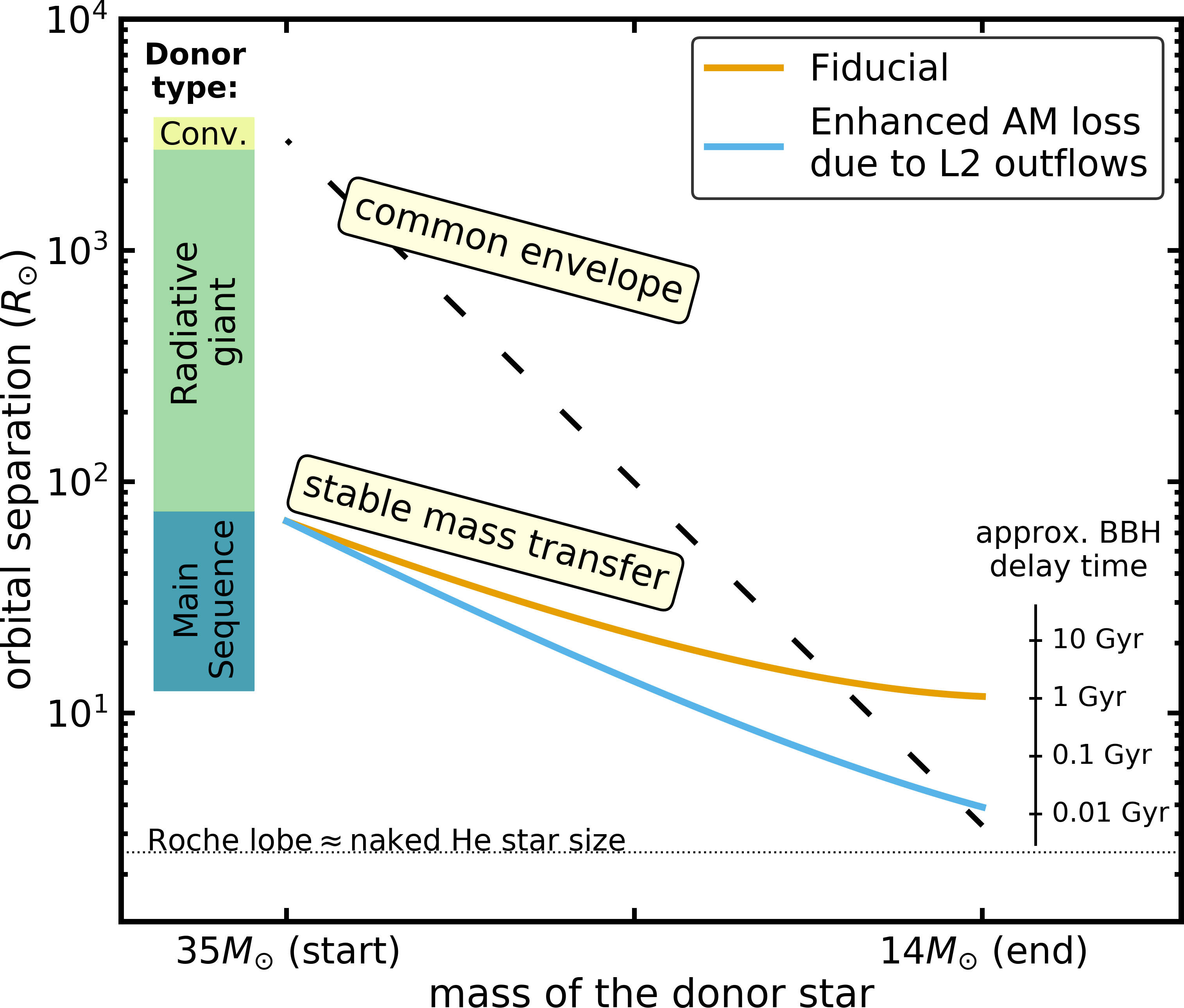}
    \caption{Schematic comparison of orbital shrinkage via common envelope and stable mass transfer in a BBH merger progenitor (here a binary with a $35\Msun$ stellar donor and $10\Msun$ BH accretor), based on a simple analytical model (Sec.~\ref{subsec.method_analytical_SMT}) rather than a detailed evolutionary calculation. The SMT lines follow assumptions typically used in population synthesis of GW sources, naively suggesting that mass transfer with enhanced angular momentum loss (blue) can produce separation as small as CE evolution. In this paper, we use detailed binary evolution calculations to show why this is not the case (Fig.~\ref{fig.fig3_dodgerblue}).} 
    \label{fig.analyt_MT_sketch}
\end{figure}

All the formation scenarios necessarily need a way to bring two BHs or NSs close together such that they spiral-in due to GW emission and merge within the age of the Universe. In isolated binaries, with the exception of chemically-homogeneous evolution \citep{Mandel2016,deMink2016,Marchant2016,Riley2021}, this is achieved through mass transfer interactions between the binary components. Classically, the key stage in the binary channel has been a phase of dynamically unstable mass transfer, leading to common-envelope (CE) inspiral \citep{Tutukov1993,vdHeuvel1994,TaurisvdHeveul2006}. This view emerged quite naturally: the dramatic shrinkage of the orbit during a CE phase is thought to produce a very short-period binary \citep{Ivanova2013,Ivanova2020} that comfortably falls into the separation range required for GW-induced mergers (Fig~\ref{fig.analyt_MT_sketch}). In comparison, stable mass-transfer (SMT) evolution can lead to all sorts of different outcomes, from moderate orbital shrinkage to orbital expansion and wide post-interaction systems \citep[][and references therein]{Tauris2023}.

In recent years, the CE-focused picture has been undergoing a revision driven by the emergence of systematic studies with detailed stellar and binary evolutionary models dedicated to the formation of GW sources \citep{Mennekens2014,Kruckow2016,Eldridge2016,Fragos2019,Klencki2020,Klencki2021,Marchant2021,Garcia2021}. These authors argued that the survival of a CE event from a massive BH progenitor may be a much rarer and fine-tuned event than previously thought, putting in question the efficiency of the CE channel for the formation of massive BBH mergers. At the same time, it was pointed out that even though the orbital shrinkage from SMT evolution is not as substantial as during a CE phace, it can also produce sufficiently short-period orbits for BBH mergers if the interaction begins in a relatively narrow orbit to begin with, see Fig~\ref{fig.analyt_MT_sketch} \citep{vdHeuvel2017,Neijssel2019,Marchant2021,GallegosGarcia2021,Picco2024}.
This realization came in part following arguments that mass transfer stability increases with the stellar mass \citep{Ge2015,Ge2020,Pavlovskii2017,Olejak2021}. Hereby, the SMT channel for BBH mergers was established, with the SMT phase in question taking place after the first BH has already been formed, at the BH+Star binary stage. 

A key question for any formation scenario is what are the orbital separations/periods of double compact objects (DCOs) it predicts. This is because the size of the orbit is the primary factor determining the time it takes for a DCO to merge ($t_{\rm merge} \propto a^4$), dictating whether GW sources preferably trace star-forming galaxies ($t_{\rm merge} \lesssim 100\,$Myr) or whether they are remnants of massive stars formed in high-redshift environments \citep[$t_{\rm merge} \gtrsim 1-10\,$Gyr;][]{Chruslinska2019,Chruslinska2019b,vanSon2022a}. The orbital separation at the BH-Wolf Rayet (WR) stage (i.e. just prior to the formation of the second BH) has a strong effect on the tidal torque \citep[$\propto a^{-6}$][]{Zahn2008} and therefore the amount of angular momentum in the WR star at core-collapse \citep{MaFuller2023}. This directly affects the amount of angular momentum in the BH progenitor at the point of core-collapse, which likely determines the BH spin. Tidally-locked BH-WR systems with orbital periods as short as $\lesssim0.5-1$ day may be required to explain the observed non-zero effective spins of BBH mergers as products of binary evolution \citep{Detmers2008, Kushnir2016,Zaldarriaga2018,Belczynski2020,Bavera2020,Bavera2021,MaFuller2023}.

Despite their importance, the orbital separations of massive DCO systems are difficult to predict in binary channels. In the case of CE evolution, 3D hydrodynamic simulations cannot yet resolve the innermost region of the envelope and model the final stages of the inspiral to find the resulting orbit (e.g. \citealt{Moreno2022}, although see \citealt{LawSmith2020}). Following CE ejection, the orbit may further be altered due to an interaction with the substantial amount of circumbinary material that has not become fully unbound \citep[possibly of a disk-like structure][]{Wei2023,Gagnier2023} or an additional mass-transfer phase \citep{VignaGomez2022,Hirai2022}. With no unambiguous massive post-CE systems discovered to date \citep{Kruckow2021}, there are few observational constraints to guide the models. Most GW-source population models predict post-CE separations according to the energy budget criteria \citep{vdHeuvel1976,Webbink1984,Livio1988} with no lower limit other than the crude requirement that the naked He core must fit into its Roche lobe in the post-CE orbit. 

In the case of SMT evolution, the main source of uncertainty for orbital shrinkage is the assumption of how much orbital angular momentum (AM) is lost with the non-accreted matter: the more AM is lost, the more the separation decreases. The key mass transfer phase takes place at a high-mass X-ray binary stage. The mass transfer rates are typically highly super-Eddington \citep{King1999,Podsiadlowski2000,Podsiadlowski2002}, leading to supercritical accretion disks and, depending on the viewing angle, an ultra-luminous X-ray source (ULX) appearance \citep{Shakura1973,King2000,Poutanen2007,Lasota2016}. Most of the transferred mass likely never reaches the BH but is instead ejected from the disk in the form of jets and bipolar outflows, disk winds, or as a slow ejecta concentrated in the equatorial plane \citep{Lipunova1999,McKinney2014,Sadowski2014,Sadowski2015,Sadowski2016}. All those components have been observed in the Galactic X-ray system SS433, the prototype for these so-called microquasars \citep{Fabrika2004,Cherepashchuk2005,Begelman2006,Blundell2008,Middleton2021}. Indirectly, they have been proposed to give rise to the observed diversity in the X-ray spectra of ULXs as the viewing angle effect \citep{Kaaret2017,KingLasotaMiddleton2023}. Whereas this general pictures holds fairly well qualitatively, we know very little about how much mass and AM is lost with each of the ejecta.
Most GW-source population studies assume that all the non-accreted matter is lost through a bipolar outflow or fast wind launched close to the BH, i.e. the isotropic re-emission model. This can be viewed as a lower limit on the AM loss.\footnote{Unless the stellar wind of the donor is comparable to the highly super-Eddington mass transfer rate through L1 of $\sim 10^{-3}\Msunyr$. }
However, having a fraction of the mass leave the system as an outflow launched far away from the BH (e.g. through the L2/L3 outer Lagrangian point) would significantly enhance the loss of AM and the degree of orbital shrinkage \citep{LuWenbin2023}. 
We illustrate this in Fig~\ref{fig.analyt_MT_sketch} as a comparison between the yellow and blue lines.\footnote{Both examples of SMT evolution in Fig~\ref{fig.analyt_MT_sketch} would satisfy the mass transfer stability criteria often used in population synthesis ($\zeta_{\rm RL} < 6.5$).} The blue model ejects 25\% of the non-accreted mass through L2 outflows. As a result, it produces a BBH merger with a delay time that is $\sim100$ times shorter, comparable to the fast-merging BBH systems that would otherwise originate only from CE evolution. 
The importance of AM loss for the SMT evolution channel has been emphasized and discussed in several recent studies \citep{vanSon2022b,Willcox2023,Picco2024,Olejak2024,GallegosGarcia2024}.

The idealized example in Fig.~\ref{fig.analyt_MT_sketch}, however, does not model the  donor star and its reaction to mass loss, which is the necessary step to gauge how much mass is transferred, on what timescale, and whether the interaction remains stable. This can be achieved with detailed stellar models of BH+Star systems \citep[e.g.][]{Podsiadlowski2003,Detmers2008, Marchant2021,GallegosGarcia2021,Klencki2021,Klencki2022,Fragos2023,Misra2024}. Although numerically more expensive than rapid binary evolution simulations, binary models involving a 1D stellar structure are crucial for realistic predictions from binary channels as, at the moment, these detailed results are not at all well reproduced by rapid GW-source population models \citep{Marchant2021,GallegosGarcia2021,Klencki2022}.  

Previous studies of the SMT channel using detailed binary evolution models of BH+Star systems explored only a limited number of cases and did not investigate the impact of the AM budget uncertainty on the properties of BBH mergers \citep{Marchant2021, GallegosGarcia2021}. Here, we fill this gap by modeling BH+Star systems with the 1D stellar-evolution code MESA (Sec.~\ref{subsec.method_mesa}) in the entire parameter space of masses, periods, and mass ratios in which the SMT channel operates at low metallicity. We explore different assumptions on the orbital AM loss and the resulting shrinkage during mass transfer.  \\
We find that there is a fundamental limit to how tight binary systems can get via stable mass transfer that is robust against uncertainties in binary physics (contrary to expectations outlines in Fig.~\ref{fig.analyt_MT_sketch}): Sec.~\ref{subsec.res_rapid_vs_detail} and  Sec.~\ref{subsec.res_there_is_a_limit}. This orbital separation limit has direct implications for the properties of BBH mergers, in particular delay times and spins (Sec.~\ref{subsec.res_consequences_BBHmergers} and Sec.~\ref{sec.disc_gw_implications}). The reason for the limit lies in the stellar structure of the donor star (as opposed to in binary processes), which dictates when the mass transfer becomes unstable (Sec.~\ref{subsec.res_it_is_the_star_that_matters} and Sec.~\ref{subsec.disc_why_the_limit_exists}). We discuss which types of donor stars can lead to BBH mergers (Sec.~\ref{sec.disc_why_donors_differ}) and how GW astronomy could yield clues into stellar interiors (Sec.~\ref{sec.disc_GW_stellar_interiors}). The results from our detailed models can be applied in rapid binary codes (Sec.~\ref{sec.disc_stability_treatment}). We conclude with Sec.~\ref{sec.summary}.

\section{Methods}
\label{sec.method}

\subsection{Overview of model grids}
\label{sec.method_overview_grids}

We compute detailed binary evolution models of BH+O-star systems, i.e. binaries comprised of a BH and a massive hydrogen-rich star. 
We calculate two types of model grids: (a) given the mass of the first-formed BH ($M_{\rm BH;1}$), grids covering different initial periods and mass ratios of BH+O-star systems; (b) given the mass and radius of a donor star, grids exploring different mass ratios to determine the critical mass ratio for SMT of the donor. 
For the first type of grids, we explored BH masses $M_{\rm BH;1}=4$, $7$, $10$, $13$, $16$, $20$, $24$, $28$, $32$, $36$, and $40 \Msun$. For each $M_{\rm BH;1}$, we considered periods and mass ratios from a wide enough range to capture the entire parameter space for the formation of BBH mergers through SMT. As the range is not known a priori, for the first few $M_{\rm BH;1}$ masses we evolved BH+O-star systems for all the initial periods that lead to mass transfer (from $\sim 1$ to $1000$ days). Most of the grids were computed at metallicity $Z = 0.1\Zsun = 0.0017$. In addition, for $M_{\rm BH;1} = 10 \Msun$ we explored several different metallicities from $Z = 0.4\Zsun$ to $Z = 0.01\Zsun$.
The BH+O-star grids are summarized in Fig.~\ref{fig.appendix_4_7_10} and Fig.~\ref{fig.appendix_metallicity}.
For the second type of grids, we explored donor masses from $10$ to $100\Msun$ and two assumptions on the efficiency of semiconvection: $\alpha_{\rm sc} = 33$ (the same as in the BH+O-star grids) and $\alpha_{\rm SC} = 0.01$. 
In addition, for both types of grids we considered two variations (three for $M_{\rm BH;1} = 10\Msun$) on how much AM is carried away with the non-accreted matter during mass transfer, introduced in Sec.~\ref{subsec.method_orbit_equations}. To better understand the results of these detailed binary models and compare with rapid binary-evolution methods, in Sec.~\ref{subsec.method_analytical_SMT} we introduce a semi-analytical model to  estimate the outcome of mass transfer evolution. Our stellar evolution computations terminate at central-carbon depletion (Sec.~\ref{subsec.method_mesa}). In Sec.~\ref{subsec.method_fromMESA_to_CC}, we details our assumptions on further evolution until the core-collapse and the formation of the secondary BH.

\subsection{Orbital evolution during stable mass transfer}
\label{subsec.method_orbit_equations}

Through considerations of the orbital AM in an interacting BH+Star system with a circular orbit\footnote{See Chapter 7 of the lecture notes by Onno Pols \url{https://www.astro.ru.nl/~onnop/education/binaries_utrecht_notes/}.}, one can express the rate of the separation change as:
\begin{equation}
    \frac{\dot{a}}{a} = - 2 \frac{\dot{M}_{\rm star}}{M_{\rm star}} \Big( 1 - \beta \frac{M_{\rm star}}{M_{\rm BH}} - (1-\beta)(\gamma+0.5) \frac{M_{\rm star}}{M_{\rm star}+M_{\rm BH}} \Big) \, \, ,
\label{eq.da_a}
\end{equation}
where $a$ is the semi-major axis (separation), $M_{\rm star}$ is mass of the mass-losing star (the donor), $M_{\rm BH}$ is mass of the accreting BH (the accretor), $\beta$ is the fraction of the transferred mass that is accreted, and $\gamma$ parametrizes the specific AM of the non-accreted matter $h_{\rm loss}$ in units of the orbital AM $J_{\rm orb}$:
\begin{equation}
    h_{\rm loss} = \frac{\dot{J}_{\rm orb}}{ \dot{M}_{\rm star}+\dot{M}_{\rm BH}} = \gamma \frac{J_{\rm orb}}{M_{\rm star}+M_{\rm BH}} \, \, \, \, .
\end{equation}
For a circular orbit, $J_{\rm orb} = \sqrt{GMa} M_{\rm star} M_{\rm BH} / (M_{\rm BH}+M_{\rm star})$. The value of $\gamma$ depends on how the non-accreted matter is lost from the system.
The assumption that it is ejected from the close proximity of the BH, carrying its orbital AM, gives:
\begin{equation}
\gamma_{\rm BH} = M_{\rm star}/M_{\rm BH} \, \, \, \, .
\end{equation}
The assumption that the mass is lost through the outer Lagrangian point L2 can be approximated with:
\begin{equation}
\gamma_{\rm L2} = \Big( \frac{a_{\rm L2}}{a}  \Big)^2 \frac{(M_{\rm BH}+M_{\rm star})^2}{M_{\rm BH}M_{\rm star}} \approx 1.44 \frac{(M_{\rm BH}+M_{\rm star})^2}{M_{\rm BH}M_{\rm star}}
\label{eq.gamma_L2}
\end{equation}
where $a_{\rm L2}$ is the distance from the L2 to the center of mass. Its value slowly depends on the mass ratio and can be approximated as $a_{\rm L2} \approx 1.2 a$ \citep{Pribulla1998}. The L2 point is located on the outer side of the BH for mass ratios $M_{\rm star} > M_{\rm BH}$ (the usual case throughout the paper), with the L3 point then located on the outer side of the donor. For mass ratios $M_{\rm star} < M_{\rm BH}$, the location of the L2 and L3 points becomes swapped. Here, for simplicity, we only consider the mass loss through L2, regardless of the mass ratio. This is justified because, as we will show, in BBH merger progenitors the mass ratio $M_{\rm star} / M_{\rm BH}$ never becomes lower than $\sim1/3$, in which case $\gamma_{\rm L2}$ does not deviate from $\gamma_{\rm L3}$ by more than 25\%. 

Finally, mass lost with the wind of the donor corresponds to $\gamma_{\rm star} = M_{\rm BH}/M_{\rm star}$. Mass loss with the specific AM of the entire orbit would correspond to $\gamma_{\rm orbit} = 1.0$. Equation~\ref{eq.da_a} only considers the AM associated with the orbit, neglecting the internal AM from the spin of the star or the BH because typically $J_{\rm spin;BH} << J_{\rm spin;star} << J_{\rm orb}$.\footnote{This is because massive stars are centrally condensed and have low moments of inertia. An exception are early MS stars interacting in short-period orbits, in which case $J_{\rm spin;star}$ may be of the order of $\sim0.1J_{\rm orb}$.} The $J_{\rm spin;star}$ term is accounted for in our detailed binary models (Sec.~\ref{subsec.method_mesa}).

As the orbit is expected to stay (nearly) circular during an SMT phase, the loss of orbital AM directly affects the orbital separation ($J_{\rm orb}^2 \propto a$) and the larger the $\gamma$, the more the orbit shrinks or the less it expands. 
 For BH-star systems with $M_{\rm star}>M_{\rm BH}$ we have $\gamma_{\rm star} <  \gamma_{\rm orbit} < \gamma_{\rm BH} < \gamma_{\rm L2}$. This is why mass loss through stellar winds ($\gamma_{\rm star}$) expands the orbit, non-conservative mass transfer with mass loss from the proximity of the BH ($\gamma_{\rm BH}$) shrinks the orbit\footnote{What is meant here is mass ejected at distance $r<<a$ from the BH, with velocity $V>>V_{\rm orb}$. This includes winds launched at the spherization radius of a supercritical disk \citep{Shakura1973}.}, and any L2 outflows ($\gamma_{\rm L2}$) enhance the orbital shrinkage even more. 
 
 Throughout the paper, we will consider the case where $\gamma$ can be divided into three components:
 \begin{equation}
     \gamma = f_{\rm wind} \gamma_{\rm star} + f_{\rm bipolar} \gamma_{\rm BH} + f_{\rm L2} \gamma_{\rm L2} \, \, \, ,
\label{eq.gamma}
 \end{equation}
where $f_{\rm wind} + f_{\rm bipolar} + f_{\rm L2} = 1$. In this case a fraction $f_{\rm bipolar}$ of the non-accreted mass is ejected from near the BH (as a bipolar outflow/jet or via fast disk wind), fraction $f_{\rm L2}$ is lost from near the L2 point, and fraction $f_{\rm wind}$ is lost from the system with the wind of the donor star. Most of the time, we will have $f_{\rm wind} << 1$ (i.e. the wind is negligible). We explore three scenarios: 
 \begin{itemize}
     \item \textbf{Fiducial treatment: $\gamma \approx \gamma_{\rm BH}$} \\
     i.e. mass ejected from the proximity of the BH with no L2 outflows.
     
     \item \textbf{Enhanced orbital shrinkage: $\gamma \approx 0.8 \gamma_{\rm BH} + 0.2 \gamma_{\rm L2}$}\\
     i.e. $\sim20\%$ of mass ejected through L2.
     
     \item \textbf{Extreme orbital shrinkage: $\gamma \approx 0.45 \gamma_{\rm BH} + 0.55 \gamma_{\rm L2}$}\\
     i.e. $\sim55\%$ of mass ejected through L2.
 \end{itemize}
 The exact fractions of mass lost through L2 in the medium/high AM loss models depend on the mass ratio of the binary, as described in detail in App.~\ref{app.L2_mishap}.

\subsection{Semi-analytical model for a mass transfer phase}
\label{subsec.method_analytical_SMT}

To better understand the results obtained in detailed binary models, we develop a semi-analytical model to calculate the outcome of an SMT phase. This method is similar to the way in which the SMT is treated in rapid binary-evolution codes widely used for GW-source population synthesis. The main two differences with respect to MESA binary models are: (a) mass transfer stability, which in MESA is obtained self-consistently for each individual binary while rapid codes rely on prescriptions, and (b) the effect of mass transfer on the donor's core mass \citep{Schurmann2024, Shikauchi2024}.

We consider a binary comprised of a BH and a massive hydrogen-rich star ($M_{\rm O;Star}$), that we will refer to as BH+O-star system. The star initiates an SMT phase when the orbital separation is $a_{\rm ini}$. For simplicity, we assume circular orbits. The product of this interaction in a binary with a BH and a massive He-rich star ($M_{\rm He;Star}$): a BH+He-star system. The final separation is:
\begin{equation}
    a_{\rm fin} = a_{\rm ini} + \int_{M_{\rm O;star}}^{M_{\rm He;star}} \frac{{\rm d}a}{{\rm d}M_{\rm star}} \left( M_{\rm BH}, \beta, \gamma \right)    {\rm d}M_{\rm star} \, \, ,
\label{eq.SMT_integral}
\end{equation}
where the expression for ${\rm d}a / {\rm d}M_{\rm star}$ can be found by combining Eqn.~\ref{eq.da_a} and Eqn.~\ref{eq.gamma}.
To compute the integral in Eqn.~\ref{eq.SMT_integral}, one needs to know how much mass was lost from the star ($M_{\rm He;star}$), whether via wind or mass transfer, as well as what were the accretion efficiency ($\beta$) and the specific AM of the non-accreted matter ($\gamma$). We describe our set of assumptions further in App.~\ref{app.semianalytical_model}.

\subsection{Detailed MESA binary models}
\label{subsec.method_mesa}

We employed the MESA stellar evolution code \citep{Paxton2011,Paxton2013,Paxton2015,Paxton2018,Paxton2019,Jermyn2023}
\footnote{MESA version r15140, \url{http://mesa.sourceforge.net/}}.
We modeled the evolution of BH+O-star systems and begin our simulations at the zero-age MS (ZAMS) of the stellar component. We thus assumed that the companion evolves like a normal, single star (in reality it had most likely been a mass gainer in a past mass transfer prior the first BH formation). The evolution was followed throughout the entire mass transfer phase (or until the mass transfer becomes unstable) and later until the central carbon depletion. 

We applied the mixing-length theory \citep{BohmVitense1958} with mixing length $\alpha = 1.5$, the Ledoux criterion for convection, and semiconvective mixing with an efficiency of $\alpha_{\rm SC} = 33$ \citep{Schootemeijer2019}.\footnote{We did not apply the convective premixing scheme introduced in \citet{Paxton2019}.} We accounted for convective core-overshooting during H- and He-burning with a step overshooting length of $\sigma_{\rm ov} = 0.33$ in the fiducial model \citep[as calibrated by][]{Brott2011}. To help converge models approaching the Eddington limit, we applied reduction of superadiabacity in radiation-dominated regions using the new implicit method from \citet{Jermyn2023}.\footnote{Notably, this does not affect the evolution of stellar radii nearly as much as the previous MLT++ treatment, see Sec.~7.2 in \citet{Jermyn2023} for comparison.}
Stellar winds were modeled following the approach of \citet{Klencki2022}, comprised of various recipes for H-rich stars \citep{Nieuwenhuijzen1990,Vink2001} and He-rich stars \citep{Nugis2000,Hainich2014,Tramper2016,Yoon2017}. In addition, based on \citet{Grafener2008}, we enhanced the mass loss rates for H-rich stars from \citet{Vink2001}\footnote{On the hot side of the bi-stability jump.} for stars approaching the Eddington limit to account for the transition to optically-thick winds \citep[see also]{Vink2011,Bestenlehner2014,Sander2020a}.
We modeled rotationally-induced mixing and internal angular-momentum transport via Taylor-Spruit dynamo, Eddington-Sweet circulation, secular shear instabilities, and the Goldreich-Schubert-Fricke instability, with a combined efficiency factor $f_c = 1/30$ \citep{Heger2000,Brott2011}.
We included rotationally-enhanced mass loss as in \citet{Langer1998}. Tidal interactions follow the synchronization timescale for radiative envelopes from \citet{Hurley2002}.
The mass transfer rate was calculated following the scheme by \citet{Marchant2021} and the accretion rate was Eddington-limitted. The scheme self-consistently accounts for L2/L3 outflows when the size of the donor reaches the L2/L3 equipotential. In practice, this happens only briefly during instances of very rapid mass transfer and the amount of mass lost via L2/L3 this way is very small ($<<1\%$). Here, we model the fraction of mass ejected through L2 ($f_{\rm L2}$) as possibly being much higher, with $f_{\rm L2} = \sim0.0$, $\sim0.2$, or $\sim0.55$ (following the three AM loss models introduced in Sec.~\ref{subsec.method_orbit_equations}, see also App.~\ref{app.L2_mishap} for exact $f_{\rm L2}$ values). The higher $f_{\rm L2}$ values account for the complicated gas-dynamics and mass ejections from BH accretion disks fed at a highly super-Eddington rate \citep[e.g.][]{LuWenbin2023}. Apart from L2 outflows, the rest of the non-accreted matter is ejected with the specific AM of the BH ($\gamma_{\rm BH}$, Sec.~\ref{subsec.method_orbit_equations}). Mass lost in stellar winds is lost with the specific AM of the star ($\gamma_{\rm star}$).

We assume a maximum mass transfer rate $\dot{M}_{\rm max} = 10^{-0.5}\Msunyr$, above which the mass transfer is deemed to become unstable and to soon enter a common-envelope phase. Although somewhat arbitrary, the choice of $\dot{M}_{\rm max}$ does not strongly affect our results, as we discuss in Sec.~\ref{subsec.disc_why_the_limit_exists}. For systems assumed to enter a CE phase, we calculate the envelope binding energy of the donor and estimate the CE survival based on the energy budget formalism, assuming $\alpha_{\rm CE} = 1.0$ \citep[similarly to][]{Klencki2021}.

\subsection{From the end point of MESA to BBH mergers}
\label{subsec.method_fromMESA_to_CC}

To put the BH+He-star systems obtained in our binary models in the context of BBH mergers, we assume that the massive He-star always collapses to form a second BH. To illustrate how the final orbital separations relate to BBH delay times, we make a simplifying assumption that the secondary BH forms in direct collapse, with no natal kick or mass loss (whether in baryons or in neutrinos). The only exception is that we account for pair-instability induced mass loss for the most massive stars (carbon-oxygen cores $M_{\rm CO} > 38 \Msun$) using fits from \citet{Renzo2022}, based on models by \citet{Farmer2019}. This results in the most massive BHs of around $\sim43\Msun$ \citep[but see][]{Farag2022}.
The presence of non-zero kicks of BHs or mass loss at BH formation would affect the predicted merger times, potentially allowing for a small population of fast merging BBH even from relative wide but eccentric orbits (see Sec.~\ref{sec.disc_kicks} and appendix in \citealt{Marchant2016}).
To calculate the merger time of a DCO system we follow the analytical fit from \citet{Mandel2021fit}, based on \citet{Peters1964}.

\subsection{Determining critical mass ratios and minimum orbital separations}
\label{subsec.qcrit_determine}

Using the numerical MESA setup and mass transfer stability criteria described in Sec.~\ref{subsec.method_mesa}, we determine critical mass ratios for donor stars with masses from $10$ to $100\Msun$. To this end, we first simulated a grid of 15 single-star evolutionary tracks from $10$ to $100\Msun$, at $Z = 0.1\Zsun$ metallicity (see Fig.~\ref{fig.HRD} for the HR diagram). For each stellar track, we consider the entire range of its radial expansion as the possible donor stars for mass transfer in binaries. Every 0.1 dex in radius of the single-star track (i.e. $\sim20$-$25$ times for each mass), we copy the exact stellar structure into a binary MESA model to examine its behavior as a donor. The companion is set to be a point mass and the initial orbit is set to be circular with a separation such that the copied star is about to fill its Roche lobe. We explore many different mass ratios $M_{\rm donor}$/$M_{\rm accretor}$, ranging from 1 to 12, with the goal of finding the critical mass ratio $q_{\rm crit}$: the largest $q$ for which the mass transfer remains stable. We do so in an iterative way: we begin with a sparse grid of mass ratios with $\Delta q = 2$, find the largest $q$ (lowest $q$) for which the mass transfer remains stable (becomes unstable), narrow down the search to mass ratios between those two values and decrease $\Delta q$. After 3-4 such iterations we steadily converge to $q_{\rm crit}$. We follow this procedure for three variations: fiducial SMT model with $\gamma = \gamma_{\rm BH}$, model with enhanced orbital shrinkage due to L2 outflows ($\gamma \approx 0.8 \gamma_{\rm BH} + 0.2 \gamma_{\rm L2}$, as well as model with $\gamma = \gamma_{\rm BH}$ and low efficiency of semiconvection $\alpha_{\rm SC} = 0.01$ (as opposed to $\alpha_{\rm SC} = 33$ in all the other grids). Ultimately, for each initial mass-radius of the donor, we determine $q_{\rm crit}$ with accuracy $\Delta q \pm 0.02$ for the fiducial model and $ \Delta q \pm 0.1$ for the variations with L2 outflows and low efficiency of semiconvection. We computed about 30-40 models with different $q$ values for each mass-radius pair, making it about 20,000 MESA binary models for both variations, with the results presented in Fig.~\ref{fig.triumvirate} and Fig.~\ref{fig.obywatel_gc}.

% % % % % % % % % % % % % % % % % % % % % % % % % % % % % % % % % % % % % % % % % % % % % % 
% % % % % % % % % % % % % % % % % % % % % % % % % % % % % % % % % % % % % % % % % % % % % % 

\section{Results}
\label{sec.results}

\subsection{From BH+O-star systems to BBH mergers: semi-analytical and detailed binary evolution models}
\label{subsec.res_rapid_vs_detail}

\begin{figure*}
    \includegraphics[width=\textwidth]{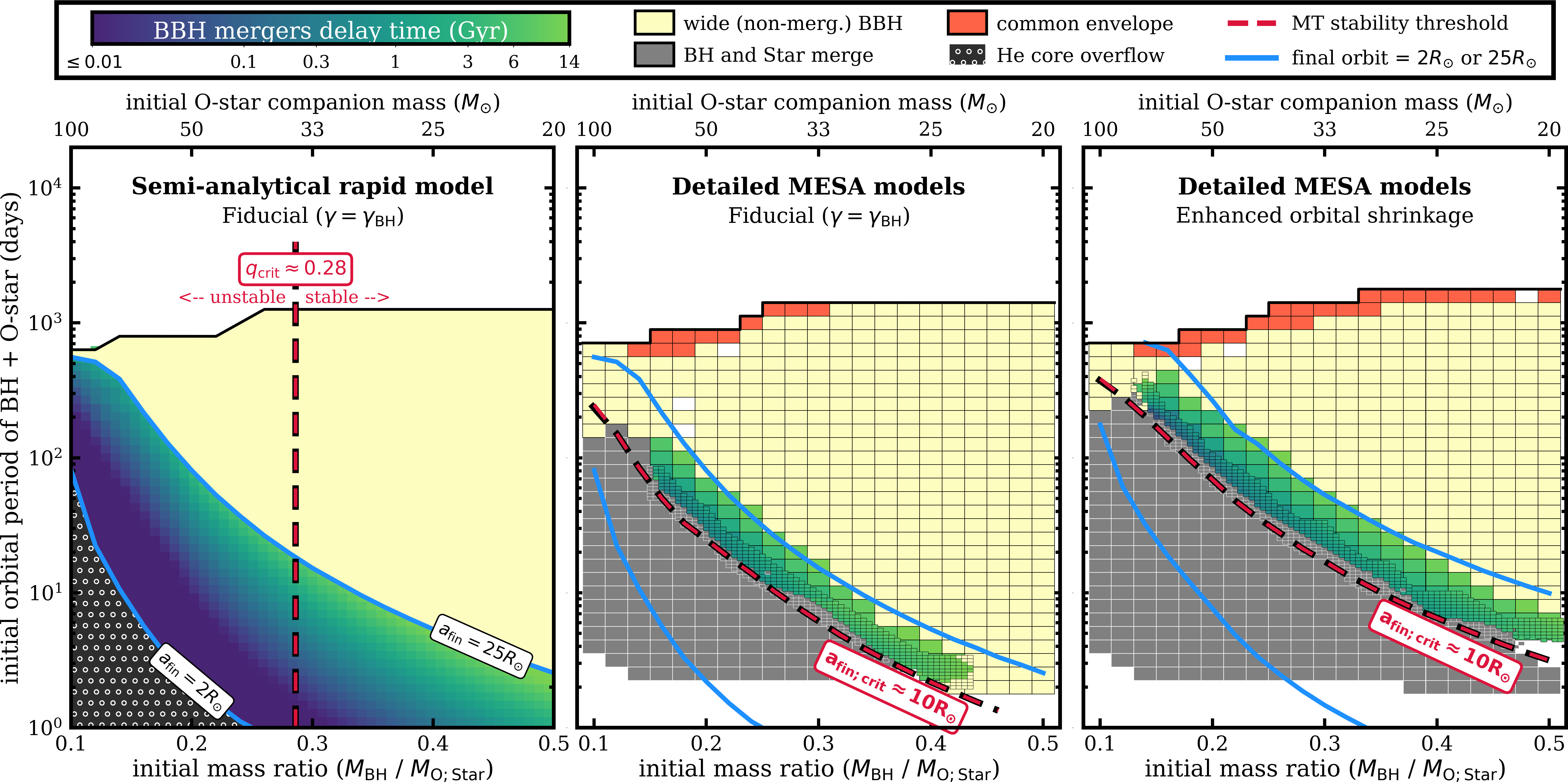}
    \caption{Window for the formation of BBH mergers, as predicted by semi-analytical and detailed binary models. Colors mark the evolutionary outcomes of interacting BH+O-star systems across different initial periods and mass ratios, calculated with our semi-analytical rapid model of SMT evolution (left panel, Sec.~\ref{subsec.method_analytical_SMT}) and compared with MESA binary models for two different assumptions on orbital shrinkage during mass transfer: fiducial ($\gamma = \gamma_{\rm BH}$, center) and enhanced shrinkage due to L2 outflows (right).    
    The initial BH mass is always $M_{\rm BH;1} = 10\Msun$ and metallicity is $Z = 0.1\Zsun$ (see Fig.~\ref{fig.appendix_4_7_10} for other BH masses). Wide-non interacting systems are not shown. 
    Solid blue lines mark systems in which the SMT evolution would lead to final orbital separation $a_{\rm fin} = 2\Rsun$ or $25\Rsun$, according to the semi-analytical model, which is approximately the range of separations needed for BBH mergers.
    The dashed red line is the boundary between stable and unstable mass transfer. The outcomes of the semi-analytical model are only applicable if the mass transfer is stable. In rapid codes, it is typically assumed to be related to a critical mass ratio (the vertical line at $q_{\rm crit} \approx 0.28$, left panel). In the MESA models, the stability is determined self-consistently at every timestep and it is found to be entirely different: the mass transfer only remains stable in systems where the final separation is $a_{\rm fin} \gtrsim 10 \Rsun$. This conclusion is robust against uncertainties of SMT evolution and holds even if enhanced orbital shrinkage is assumed (right), leading to BBH mergers with long delay times $\gtrsim 1$ Gyr (Sec.~\ref{subsec.res_consequences_BBHmergers}).}
    \label{fig.grid_analyt_vs_MESA}
\end{figure*}

Here, we analyze the evolution of BH+O-star systems with different initial orbital periods and mass ratios to investigate which of them form BBH mergers. 
To develop intuition, we begin with the rapid semi-analytical method for mass transfer evolution (Sec.~\ref{subsec.method_analytical_SMT}) in the left panel of Fig.~\ref{fig.grid_analyt_vs_MESA}, before moving to results from detailed MESA binary models. We consider BH+O-star systems with initial periods from $\sim$1 to 1000 days (Y axis; wider orbits would be non-interacting systems) and mass ratios $M_{\rm BH;1}/M_{\rm O;Star}$ from 0.1 to 0.5 (X axis). The initial BH mass is always $M_{\rm BH;1} = 10\Msun$, meaning the initial O-star mass ranges from 20 to 100 $\Msun$.
In each binary, as the O-star evolves and expands, it will eventually initiate mass transfer onto the BH. By that point, the period and mass ratio have changed slightly from the initial values due to winds. For the purpose of the semi-analytical exercise, we consider the mass transfer to be always stable and simply mark the potential stability criteria at critical mass ratio $q_{\rm crit} = \approx 0.28$ (dashed vertical line, after \citealt{Riley2022_compas} who follow \citealt{Ge2015}). For simplicity, we assume fully
non-conservative mass transfer ($\beta = 0$) as the mass transfer rates
expected in BH binaries with massive donors are generally highly
super-Eddington \citep{Podsiadlowski2003,Rappaport2005}. Indeed, in our MESA models with Eddington-limited accretion we find $\beta \lesssim 0.1$.
We assume that the non-accreted matter is ejected from near the BH ($\gamma = \gamma_{\rm BH}$). For each BH+O-star system, we calculate the final post-SMT separations $a_{\rm fin}$ according to Eqn.~\ref{eq.SMT_integral}. Assuming the secondary collapses to form the second BH (Sec.~\ref{subsec.method_fromMESA_to_CC}), we obtain the corresponding BBH merger delay times. The results are in the left panel of Fig.~\ref{fig.grid_analyt_vs_MESA}, where we color-code the parameter space of BH+O-star binaries according to the final outcome of SMT evolution. The two solid blue lines mark which BH+O-star binaries would reach the final separation $a_{\rm fin}$ = 2$\Rsun$ or 25$\Rsun$, which is approximately the range of orbits required for BBH systems to merge within the Hubble time. Importantly, both the $a_{\rm fin}$ values and the final outcomes in the left panel assume that the mass transfer is always stable, which in reality is not case. In most rapid binary codes, the stability criteria is comprised of a set of critical mass ratio values (or their corresponding critical $\zeta_{\rm crit}= {\rm d \, log} R_{\rm RL} / {\rm d \, log} M_{\rm star}$ exponents), see also Sec.~\ref{sec.disc_stability_treatment}. 
A fixed critical mass ratio for stability would mean excluding systems to the left of a vertical line in Fig.~\ref{fig.grid_analyt_vs_MESA}, as illustrated by a dashed red line example at $q_{\rm crit} = 0.286 \approx 1/3.5$ in the left panel. This choice of $q_{\rm crit}$ corresponds to the stability criteria for radiative giants in the COMPAS code \citep[assuming non-conservative mass transfer, $\gamma = \gamma_{\rm BH}$, and their $\zeta_{\rm crit} = 6.5$,][]{VignaGomez2018,Riley2022_compas}.

The area between the $a_{\rm fin}$ = 2$\Rsun$ and 25$\Rsun$ lines in the left panel of Fig.~\ref{fig.grid_analyt_vs_MESA} marks the potential parameter window for the formation of BBH mergers, colored by the BBH delay time.
The dark area on the left are systems that would technically produce BH+He-star systems with separations $< 2\Rsun$. In such orbits, even the naked He-star would overflow its Roche lobe, potentially leading to a merger and a transient \citep{Metzger2022} or reducing the He-core mass so substantially that it never forms a BH, unless the interaction can be cut short (e.g. by core collapse or the orbit widening).
Finally, the light yellow region are systems in which the SMT evolution would lead to the formation of wide non-merging BBH systems. The stability requirement of $q>q_{\rm crit}$ reduces the parameter space for BBH merger formation to initially short-period systems ($\lesssim 10$ day) that interact on the MS, i.e. case A mass transfer. Coincidentally, the treatment of case A evolution in rapid codes has been shown to suffer particularly significantly from the necessary simplifications of the rapid method, under-predicting the final core mass and over-predicting the amount of mass transferred \citep{Belczynski2022,RomeroShaw2023,Shikauchi2024,Schurmann2024}.

% \begin{figure*}[!b]
\begin{figure*}
\sidecaption
    \includegraphics[width=12cm]{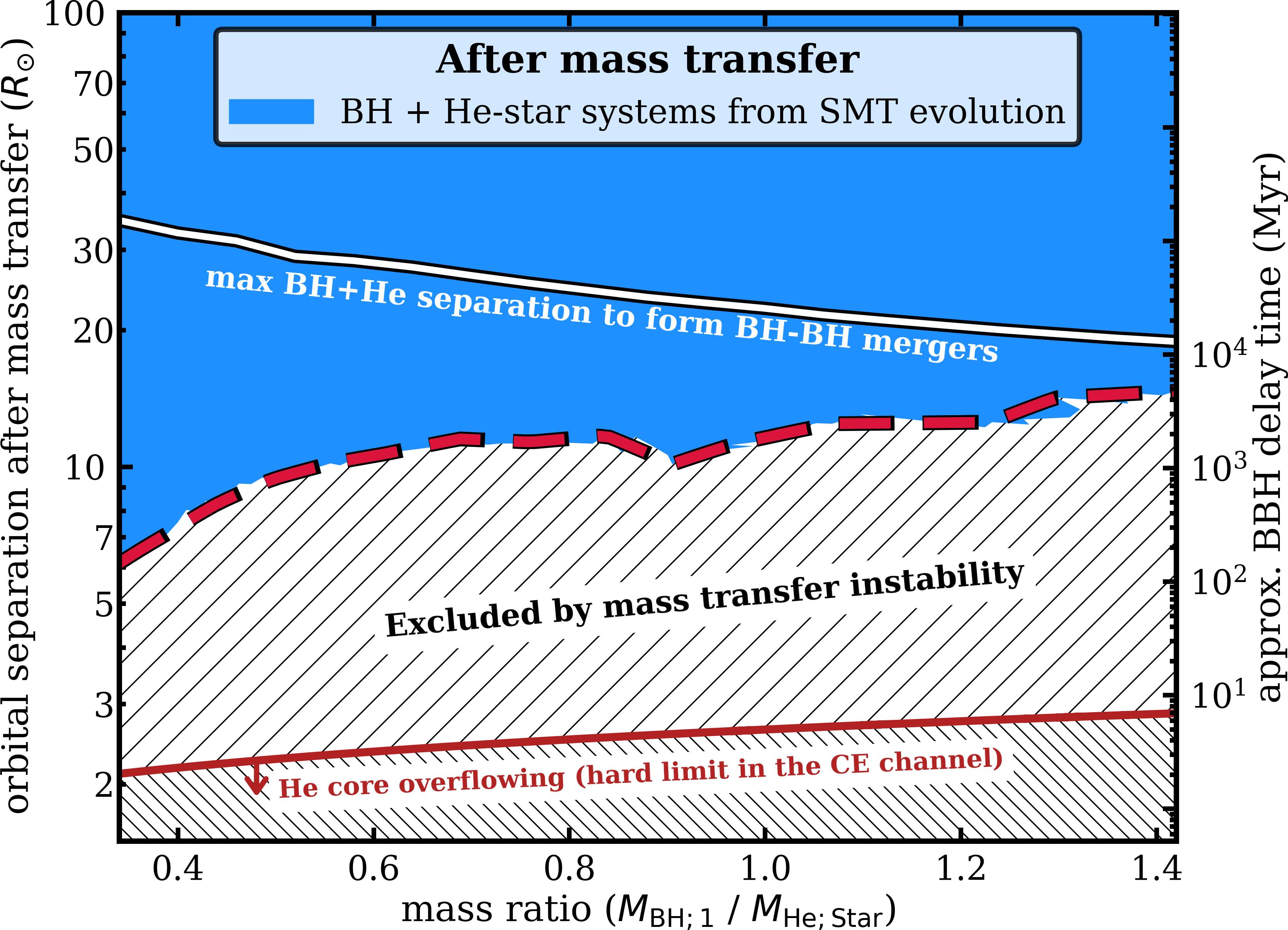}
    \caption{The fundamental orbital separation limit from stable mass transfer (red dashed line) seen in the population of BH+He-star systems formed in MESA binary models. Separation below $\sim$8-10$\Rsun$ are excluded in the SMT channel due to mass transfer instability (dashed region, physical origin in Sec.~\ref{subsec.disc_why_the_limit_exists}). As a result, BBH mergers from SMT evolution have long delay times of $\gtrsim 1$ Gyr and their BH+He-star progenitors tend to be outside the regime of tidal spin-up ($P \gtrsim 0.8$ day).
    Here, the orbital evolution during SMT assumes enhanced orbital shrinkage due to L2 outflows with $\gamma \approx 0.8 \gamma_{\rm BH} + 0.2 \gamma_{\rm L2}$. However, models assuming a different $\gamma$ yield a similar range of BH+He-star separations (Fig.~\ref{fig.before_and_after_L2}). The solid red line indicates a hard lower limit for the separation of BH+He-star systems in the CE channel, set by the size of a naked helium star. 
    The figure is based on a grid of BH+O-star models with $M_{\rm BH;1} = 10\Msun$ and $Z = 0.1\Zsun$ metallicity, with similar results found also for larger BH masses (Sec.~\ref{subsec.res_consequences_BBHmergers}). \\
    }
    \label{fig.fig3_dodgerblue}
\end{figure*}

The results from MESA binary models are presented in the other two panels of Fig.~\ref{fig.grid_analyt_vs_MESA}, for both the assumption of low AM loss during SMT (fiducial model, $\gamma = \gamma_{\rm BH}$) as well as the enhanced orbital shrinkage due to L2 outflows ($\gamma \approx 0.8 \gamma_{\rm BH} + 0.2 \gamma_{\rm L2}$). Here, we focus mainly on the final evolutionary outcomes of these models and the question whether the mass transfer has remained stable and the final separation is close enough to form a merging BBH system. We refer to \citet{Marchant2021} for a detailed overview of the evolutionary pathway from BH+O-star systems to BBH binaries, and to \citet{Podsiadlowski2002,Podsiadlowski2003} for a general discussion of mass transfer rates and interaction timescales in BH binaries.
Each rectangle in Fig.~\ref{fig.grid_analyt_vs_MESA} represents one MESA binary model. Non-interacting wide systems are not shown. Models in which the mass transfer remains stable and the second BH forms are color-coded according to the BBH delay time, with non-merging wide BBHs shown in yellow. Models in which the mass transfer becomes unstable are labeled either as 'common envelope' (for convective-envelope donors, $P_{\rm ini} \gtrsim 700$ day) or as 'BH and Star merge' (for radiative-envelope donors, $P_{\rm ini} \lesssim 700$ day). This choice is based on the expectation that unstable mass transfer and CE evolution with radiative donors lead to stellar mergers \citep{Kruckow2016,Ivanova2020,Klencki2021,Marchant2021}.\footnote{Other outcomes may be possible if, despite the mass transfer instability, a dynamical in-spiral and a classical CE phase are avoided.} The two blue solid lines mark the parameter space for BBH merger formation found in the semi-analytical model ($a_{\rm fin} = 2$ and $25\Rsun$ in the left panel), adjusted for the value of $\gamma$. Notably, the upper boundary coincides with the widest MESA models that form BBH mergers. This indicates that the orbital evolution in the semi-analytical model generally agrees well with the orbital evolution in MESA, with the exception of mass ratios $q \gtrsim 0.15$ and initial periods $P \lesssim 3$ day (for $\gamma = \gamma_{\rm BH}$). The first case is mass transfer from very massive donors ($> 70\Msun$), which in MESA transfer a lower amount of mass before detaching compared to $\Delta M_{\rm star} = 0.9 M_{\rm env}$ assumed in the rapid method, such that the orbit does not shrink as significantly. The second case, i.e. short-period systems, are binaries following early case A mass transfer evolution, which in MESA produce lower secondary BH masses compared to our rapid method due to the core mass reduction \citep{Schurmann2024}.

The striking difference between the rapid and detailed binary models is in the mass transfer stability (red dashed lines in Fig.~\ref{fig.grid_analyt_vs_MESA}).
In MESA calculations, the stability threshold for radiative-envelope donors ($P \lesssim 700^{\rm d}$) is not at all a vertical line that would correspond to any critical mass ratio $q_{\rm crit}$, or even a combination of several $q_{\rm crit}$ values. Instead, the stability line spans a wide range of mass ratios and depends on the size of the orbit: the longer the orbital period, the more stable the mass transfer in BH+Star systems, see also \citep{Marchant2021,GallegosGarcia2021}. This result agrees with the findings of \citet{Ge2015,Ge2020,Ge2024} who find that critical mass ratios vary significantly as massive stars expand, such that interactions in wider orbits (with more expanded radiative giants) are more stable. 

The mass transfer (in)stability found in detailed models directly affects the properties of BBH mergers. Most of the potential parameter space for BBH merger formation is excluded due to instability (BH + Star merge). The BBH mergers that do form in MESA have relatively wide final separations $a_{\rm fin}$ between $\sim10$ and $25\Rsun$ (for $M_{\rm BH;1} = 10\Msun$), leading to long delay times $\gtrsim 1$ Gyr. Surprisingly, this is also the case in the variation with enhanced AM loss due to L2 outflows (right panel). Even though the orbits shrink significantly more and BBH mergers may originate from initially wider BH+O-star systems, no BBH mergers with $a_{\rm fin} \lesssim 10\Rsun$ and short delay times are produced. In the following sections, we explain this as the fundamental separation limit of the SMT evolution.

\subsection{There is a limit: minimum orbital size from SMT evolution}
\label{subsec.res_there_is_a_limit}

\begin{figure*}
    \includegraphics[width=\textwidth]{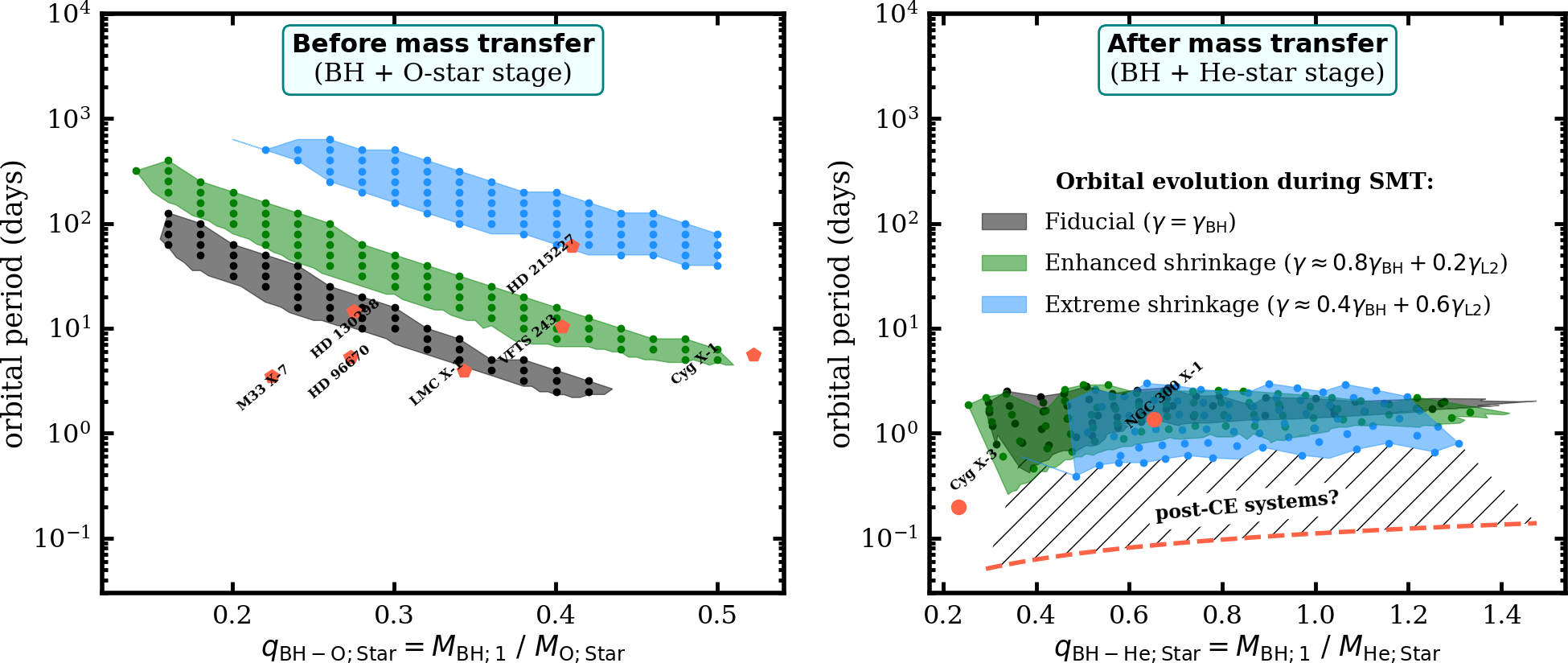}
    \caption{The narrow window in the parameter space for the SMT channel converges to a well-defined, narrow range of final orbits, regardless of uncertainties in binary physics.
    We plot the orbital periods of BBH merger progenitors from before the mass transfer (BH+O-star stage, left) and after the SMT phase has taken place (BH+He-star stage, right). The figure is based on grids of detailed binary models with $M_{\rm BH;1} = 10\Msun$ and varying assumptions on the orbital shrinkage during mass transfer (i.e. $\gamma$ values, shown in different colors). Changing the assumption on $\gamma$ significantly shifts the pre-SMT periods of BBH merger progenitors, but has little effect on the post-interaction orbits at the BH+He-star stage (or the properties of BBH mergers). The dashed red line shows the lower limit on the period of a BH+He-star systems, such that the naked helium star does not overflow its Roche lobe. We mark the periods and mass ratios of the known BH+OB-star systems (left panel, including supergiant high mass X-ray binaries) and of BH+Wolf-Rayet X-ray binaries (right). Cyg X-3 cannot be reproduced by SMT in our models, suggesting its origin in CE evolution. 
    }
    \label{fig.before_and_after_L2}
\end{figure*}

We find that there is limit on how close binary systems can get via mass transfer without triggering dynamical instability, illustrated in Fig.~\ref{fig.fig3_dodgerblue}. 
The figure is based on a grid of detailed MESA binary models of BH+O-star systems with $M_{\rm BH;1} = 10 \Msun$ (same as the right-hand panel of Fig.~\ref{fig.grid_analyt_vs_MESA}), with similar results found for other BH masses.
The hatched region marks the range of separations that we find to be excluded in the SMT channel due to mass transfer instability, with the densely hatched region also excluded in the CE channel.
% We discuss the physical mechanism of this instability in Sec.~\ref{subsec.disc_why_the_limit_exists}. 
Figure~\ref{fig.fig3_dodgerblue} demonstrates that the minimum orbital size of BH+He-star systems from SMT evolution is around $\sim 8-10\Rsun$ (periods $\sim 0.8$ day). 
This has consequences for the properties of BBH mergers: long delay times ($\gtrsim 1$ Gyr) and inefficient tidal spin-up at the BH+He-star stage (Sec.~\ref{subsec.res_consequences_BBHmergers} and Sec.~\ref{sec.disc_gw_implications}).
More broadly, the separation limit from SMT evolution depends on the type of the donor star (Sec.~\ref{subsec.res_it_is_the_star_that_matters}).We note a trend between the minimum separation of BH+He-star systems and their mass ratio $q_{\rm BH-He;Star} = M_{\rm BH;1} / M_{\rm He;Star}$ in Fig.~\ref{fig.fig3_dodgerblue}. This is because systems with lower $q_{\rm BH-He;Star}$ (i.e. more massive He stars) originate from wider BH+O-star systems, which undergo case B/C mass transfer with donors that are core-He burning stars. We find that mass transfer from such donors may remain stable in tighter orbits than case A interaction.

\begin{figure*}
    \includegraphics[width=\textwidth]{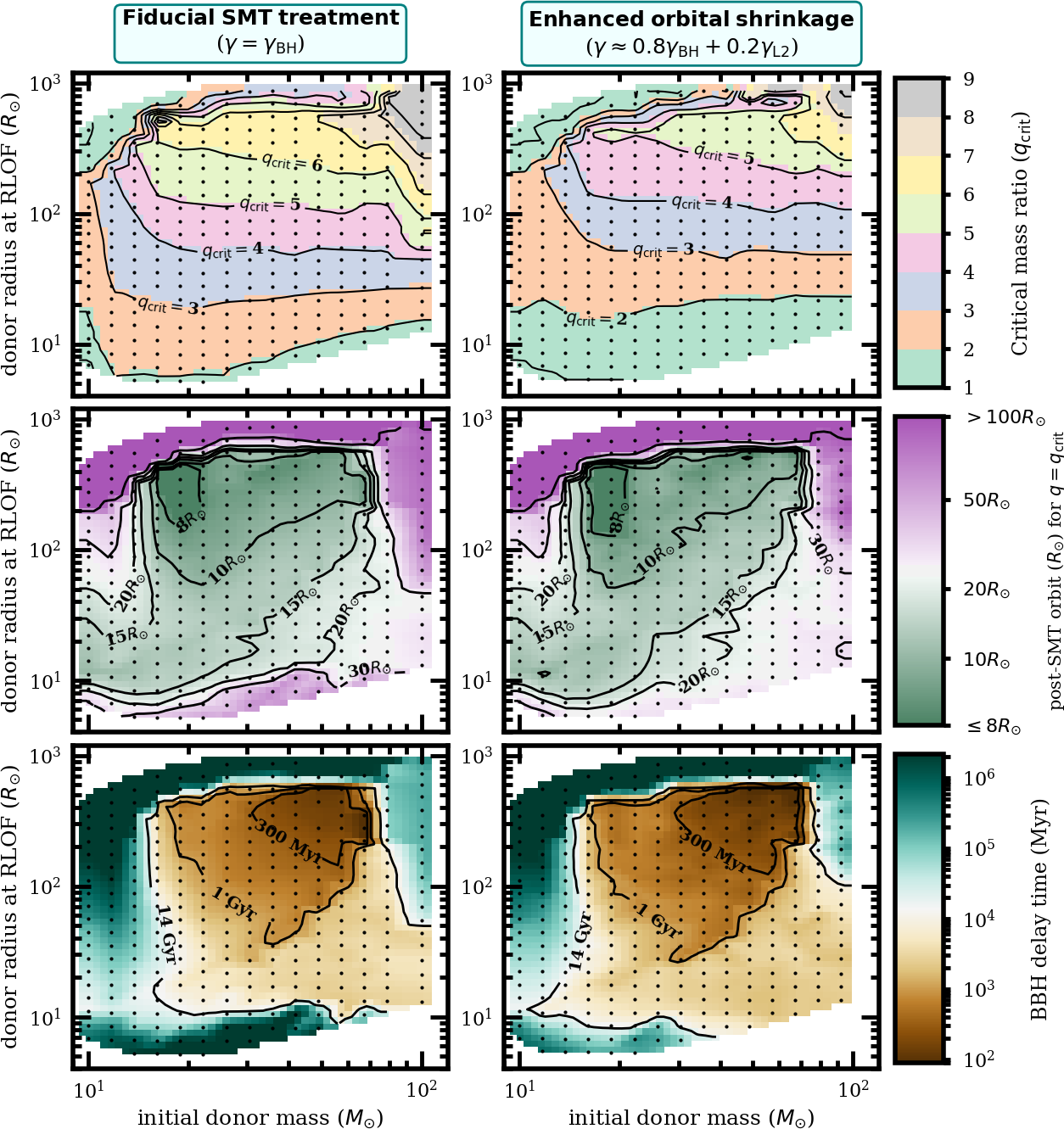}
    \caption{The minimum orbital size from SMT evolution is primarily determined by the donor star, rather than by binary physics. Here, we report the critical mass ratios $q_{\rm crit}$ for BH+O-star systems (top panels), final orbital separations in binaries with $q_{\rm RLOF} = q_{\rm crit}$ (middle panels), and their corresponding BBH delay times (bottom panels), calculated with MESA for a grid of donor stars of different mass (at ZAMS) and radius (at the onset of RLOF) at $Z = 0.1\Zsun$. 
    The separations and BBH delay times shown here can be viewed as the minimum values achievable from SMT evolution given the donor mass and radius. Left panels follow the fiducial SMT treatment with $\gamma = \gamma_{\rm BH}$, right panels assume enhanced orbital shrinkage due to L2 outflows. While enhancing the orbital shrinkage (left vs right) affects the critical mass ratios, the post-SMT orbits and their corresponding BBH delay times remain nearly unchanged. This is because these final outcomes are primarily determined by the stellar structure of the donor, which dictates when the mass transfer becomes unstable. 
    For illustrative purposes, the BBH delay times assume that each donor collapses to form the second BH, no matter its final core mass. Only the systems with delay times below $\sim 13.7$ Gyr (i.e. brown in the bottom panels) are potential BBH mergers and GW sources for the LIGO/Virgo/Kagra detectors.  
    The scatter points indicate the explored grid of donor masses and radii. For each pair, the $q_{\rm crit}$ value was found iteratively via binary MESA models down to at least $\pm 0.1$ accuracy.}
    \label{fig.triumvirate}
\end{figure*}

Counter-intuitively, the separations of BH+He-star systems from SMT evolution only weakly depend on the degree of orbital shrinkage and the choice of $\gamma$. In Fig.~\ref{fig.before_and_after_L2}, we compare the range of periods of BBH merger progenitors before the SMT phase (BH+O-star) and after the interaction (BH+He-star) for three different assumptions on the amount of L2 outflows and the corresponding AM loss.
The more mass is lost through L2, the more the orbit shrinks during SMT, meaning that wider BH+O-star systems evolve to become BBH mergers (left panel). This difference can be quite dramatic, shifting BBH merger progenitors from BH+O-star systems on a few-day period orbits (case A mass transfer) to wide dormant BH binaries with periods of hundreds of days, essentially covering the entire parameter space for interaction with radiative stars \citep{Klencki2020}. The uncertainty in $\gamma$ causes a difficulty in predicting which of the observed BH-Star systems are progenitors of GW sources. For comparison, we mark the sample of BH high-mass X-ray binaries and X-ray quiet massive BH-Star systems with known periods and mass ratios: Cyg X-1 \citep{MillerJones2021Sci},  LMC X-1 \citep{Orosz2009}, M33 X-7 (\citealt{Orosz2007}, although see \citealt{Ramachandran2022_M33}), HD 130298 \citep{Mahy2022}, VFTS 243 \citep{Shenar2022}, HD 96670 \citep{GomezGrindlay2021}, and HD 215227 \citep{Casares2014Natur}, although see \citep{Janssens2023}. 

At the BH+He-star stage, on the other hand, the orbits are similar across all the variations (right panel). Even with extreme orbital shrinkage, the SMT evolution does not lead to periods below $\sim0.6$ day, with most BH+He-star systems at $P \gtrsim 1$ day. As in Fig.~\ref{fig.fig3_dodgerblue}, shorter periods (closer separations) are excluded by mass transfer instability. 
The small differences in BH+He-stars systems in the three variations in Fig.~\ref{fig.before_and_after_L2} are due to differences in the structure of donor stars interacting at different $P_{\rm BH-O;Star}$ periods, rather than due to the mechanism of orbital shrinkage or the amount of AM loss. As we discuss in the following: it is the star that matters.

\subsection{The inside matters: which donor stars lead to short-period systems and BBH mergers?}
\label{subsec.res_it_is_the_star_that_matters}

It is not a coincidence that mass transfer becomes unstable in systems heading to orbital separations below $\sim 8 \Rsun$.
By examining individual cases (Sec.~\ref{subsec.disc_why_the_limit_exists}), we find that in such binaries, a delayed dynamical instability is triggered when the near-core envelope layers of the donor star are being transferred. These layers are located at radial coordinates $\sim3-10\Rsun$ inside the radiative envelope, depending the stellar mass and evolutionary stage. They are characterized by a particularly fast rate of expansion in response to mass loss compared to the rest of the envelope due to their near-flat entropy profile. 
To keep the mass transfer stable, the orbit cannot continue to shrink significantly anymore when the flat-entropy layers are being stripped. 
This prevents the formation of BH+He-star systems with separations $\lesssim 8-10 \Rsun$ via SMT evolution, regardless of the overall degree of orbital shrinkage from the moment of RLOF and the choice of $\gamma$.
We discuss the mechanism of near-core instability in more detail in Sec.~\ref{subsec.disc_why_the_limit_exists}. 

This signals an important point: 
\textit{orbits of BH+He-star systems formed in the SMT channel depend primarily on the donor stars and their inner entropy profiles rather than on the AM loss or accretion efficiency during the SMT phase.} \\
The internal structure of stars, on the other hand, can vary greatly depending on the stellar mass and radius, evolutionary stage, and chemical mixing. 
To examine which types of donors favor the formation of close BH+He-star systems and BBH mergers, we systematically compute mass transfer evolution from stars of different mass (from $10$ to $100\Msun$) and size (the entire range of radial expansion) at $Z = 0.1\Zsun$. For each mass-radius pair of the donor, we determine the critical mass ratio for SMT evolution $q_{\rm crit}$ following methodology described in Sec.~\ref{subsec.qcrit_determine}.
The result is shown in the upper panels of Fig.~\ref{fig.triumvirate} for two variations: with fiducial SMT treatment ($\gamma = \gamma_{\rm BH}$, left panel) and with enhanced orbital shrinkage ($\gamma \approx 0.8 \gamma_{\rm BH} + 0.2 \gamma_{\rm L2}$, right panel).

The $q_{\rm crit}$ values we find in Fig.~\ref{fig.triumvirate} have several characteristics. For a given donor mass, $q_{\rm crit}$ increases as a function of the donor radius all the way until the donor develops an outer convective envelopes ($\gtrsim200$-$700\Rsun$, depending on the mass). At that point, $q_{\rm crit}$ rapidly decreases. We also see a slow trend with mass: the more massive a donor, the larger its $q_{\rm crit}$ values. This is consistent with the prior large-scale studies of mass transfer stability, which also discuss the origin of those trends in detail \citep{Ge2015,Ge2020,Ge2024}. Notably, both these past as well as our results indicate that an assumption of a single $q_{\rm crit}$ value for radiative or convective donors is an oversimplification, as the trend with radius is strong. Finally, we observe that all the $q_{\rm crit}$ values decrease when enhanced orbital shrinkage is assumed (right vs left in Fig.~\ref{fig.triumvirate}). This is fully expected: the more rapidly the orbit shrinks, the more unstable the mass transfer is. While it is possible to analytically compute correcting factors to $q_{\rm crit}$ given any chosen $\gamma$ such that the rate of shrinkage at the onset of RLOF is the same\footnote{Meaning the $\zeta$ exponent stays the same, where $\zeta= {\rm d \, log} R_{\rm RL} / {\rm d \, log} M_{\rm donor}$.}, these would not be correct, because the instability is not determined at the moment of RLOF but at some later stage, a priori unknown.

In the central panels of Fig.~\ref{fig.triumvirate}, we show the orbital separations at the end of SMT in binary models with $q = q_{\rm crit}$ as a function of the initial donor mass and its radius at RLOF. These can be viewed as the minimum orbital size of BH+He-star system that may be formed as a result of SMT evolution, given the donor star.\footnote{In a system with $q$ smaller than $q_{\rm crit}$, the orbit would shrink less during the phase when $M_{\rm donor} > M_{\rm accretor}$ and it would expand more once/if the mass ratio becomes reversed ($M_{\rm accretor} > M_{\rm donor}$).} As we are interested in close-orbit BH+He-star systems, we focus the color-scale on the range of separation from $8$ to $100 \Rsun$, without resolving the 
wider orbits of systems that underwent SMT with convective donors. Assuming that all the He-stars collapse to form the secondary BHs (Sec.~\ref{subsec.method_fromMESA_to_CC}), we further compute the delay times of the resulting BBH systems and color-code the result in the bottom panels of Fig.~\ref{fig.triumvirate}. Delay times $\lesssim 13.7$ Gyr (brown) are BBH systems that merge within the age of the Universe. 
We find that donors with masses $\lesssim 20\Msun$ do not lead to BBH mergers, which may set the lower mass limit for the SMT channel.

\begin{figure*}
    \includegraphics[width=\textwidth]{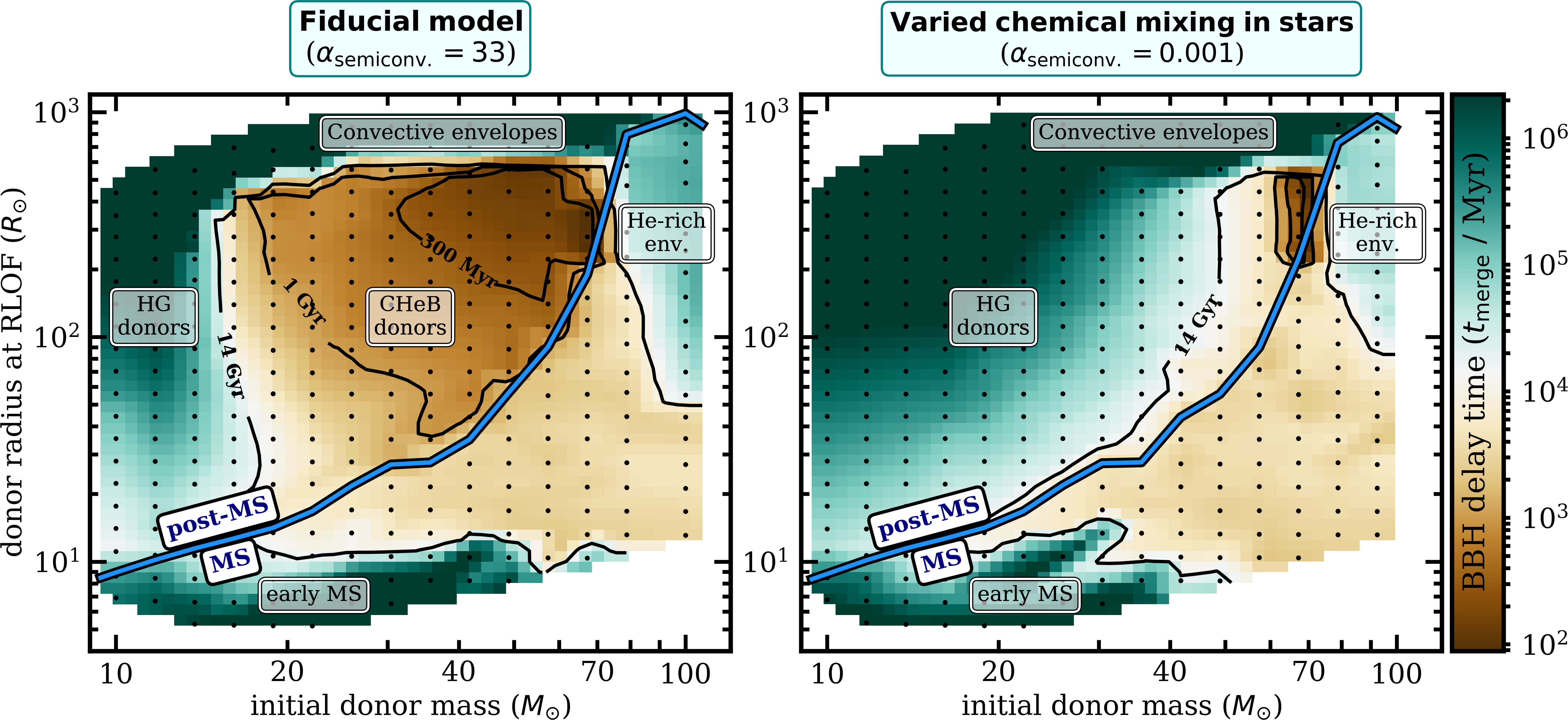}
    \caption{The formation of BBH mergers is a probe of internal chemical mixing in stars and of the stellar structure. We plot the minimum BBH delay times from SMT evolution given the donor mass (at ZAMS) and radius (at RLOF) for two variations in the efficiency of semiconvective mixing in stars ($\alpha_{\rm semiconv.}$). 
    Only cases with the BBH delay time below $\lesssim 13.7$ Gyr (yellow-brown regions) are potential GW sources. The range of donors that may lead to BBH mergers is severely reduced in the variation with low efficiency of semiconvection (right panel) and it shows a preference for high-mass donors ($\gtrsim 40\Msun$). This is because the chemical structure of donor stars directly affects mass transfer stability and therefore the minimum post-SMT separations and BBH delay times (Sec.~\ref{sec.disc_why_donors_differ}). The difference between stellar models in both panels is closely connected to the blue-supergiant problem (Sec.~\ref{sec.disc_GW_stellar_interiors}).
    The thick blue line shows the transition from MS to post-MS donors. 
    Several main types of donors are labeled (Hertzsprung gap, core-He burning, convective envelopes, He-rich envelopers), see text.
    The figure is based on MESA models of BH+O-star binaries with donors from $10$ to $100\Msun$, evolving through SMT at the critical mass ratio $q = q_{\rm crit}$.
    See Fig.~\ref{fig.appendix_metallicity} for the post-SMT separations.}
    \label{fig.obywatel_gc}
\end{figure*}

Fig.~\ref{fig.triumvirate} reveals a surprising result: whereas the critical mass ratios change significantly when varying the assumption on $\gamma$ (left vs right), the minimum orbital separations and BBH delay times from SMT evolution remain virtually unchanged. 
In other words, BH+He-star orbits in the SMT channel appear to be robust against the major uncertainties of orbital evolution during SMT, which is a rare case in binary astrophysics. It also suggest that for radiative donors, the physical origin of instability is more closely related to the orbital evolution at advanced stages of mass transfer (and therefore the final orbit), rather than to the initial rate of orbital shrinkage at RLOF. We discuss this in the context of delayed-dynamical instability and stability criteria in Sec.~\ref{subsec.disc_why_the_limit_exists}. 

In addition, the post-SMT orbital separations in Fig.~\ref{fig.triumvirate} show a complex behavior rather than a simple trend with mass or radius. The corresponding BBH delay times indicate that for some donor stars, the SMT channel cannot lead to sufficiently close orbits to produce BBH mergers (i.e. $t_{\rm delay} > 14$ Gyr), even for the most favorable case of $q = q_{\rm crit}$.
It turns out that the donors that 'do not work' in the SMT channel can be identified as several distinct types of stars and linked to evolutionary stages and envelope characteristics. 
We do so in Fig.~\ref{fig.obywatel_gc}, where we re-plot in more detail the minimum BBH delay times (for $q = q_{\rm crit}$) obtained from different donors
in our Fiducial model (left panel) and compare with a model variation with a severely reduced efficiency of semiconvective mixing $\alpha_{\rm semiconv.} = 0.001$ (right panel). 
We choose to vary semiconvection as it is a known factor to affect the evolutionary types of giant donors in 1D stellar models \citep{Schootemeijer2019,Klencki2020,Kaiser2020}.\footnote{The critical mass ratios $q_{\rm crit}$ in the $\alpha_{\rm semiconv.} = 0.001$ variation were computed in the same as in the other variations.} The thick blue line in Fig.~\ref{fig.obywatel_gc} marks the boundary between MS and post-MS donors. 

The largest donors with $R \gtrsim 500-700\Rsun$ are red supergiants with outer convective envelopes. They are the most prone to unstable mass transfer, with low $q_{\rm crit}$ values. The SMT evolution in their case leads to wide BH+He-star orbits of hundreds of $\Rsun$ and delay times $>> 14$ Gyr. Instead of SMT, these donors may lead to NS or BH mergers via CE evolution \citep{Kruckow2016,Klencki2021}.

The other post-MS stars are radiative-envelope giants. They can be split into two categories \citep{Klencki2020}: Hertzsprung Gap (HG) donors that engage into mass transfer right after the end of MS, as a result of a rapid envelope expansion when the core contracts, and Core-He burning (CHeB) donors that initiate mass transfer at a more advanced evolutionary stage, as a result of a slow nuclear-timescale expansion during the core-He burning phase. We find that this distinction plays a crucial role in the SMT channel. As it turns out, all the post-MS donors that cannot produce BBH mergers ($t_{\rm merge} > 14$ Gyr in Fig.~\ref{fig.obywatel_gc}) are HG donors. On the other hand, CHeB donors lead to delay times as short as $\sim 300$ Myr. 
As we discuss in Sec.~\ref{subsec.disc_why_the_limit_exists}, the reason lies in the different inner entropy structure of HG and CHeB stars. As a result, the mass transfer from HG donors leads to higher mass transfer rates and is less stable than from CHeB donors. Even the smallest orbital separations of BH+He-star systems from HG donors are comparatively quite large ($\gtrsim 20-30\Rsun$), whereas CHeB donors may lead to systems with $\sim 8\Rsun$ orbits (Fig.~\ref{fig.triumvirate}). 

The significance of the HG vs CHeB distinction becomes apparent when comparing both panels of Fig.~\ref{fig.obywatel_gc}. In the variation with reduced efficiency of semiconvection (right), all massive stars expand rapidly at the end of the MS and consequently all the radiative post-MS donors are of the HG type \citep[][]{Schootemeijer2019,Klencki2020}. They do not lead to the formation of BBH mergers via the SMT channel, nearly completely quenching the role of post-MS mass transfer. It is not a surprise that a variation in assumptions behind 1D stellar models lead to somewhat different outcomes, but it is rare for an uncertainty in the inner structure of stars to have such a strong effect on GW-source predictions. Whether massive stars interact as HG or CHeB donors is closely related to the long-standing debate on the nature of blue supergiants, as we discuss in Sec.~\ref{sec.disc_GW_stellar_interiors}.

\begin{figure*}
    \includegraphics[width=\textwidth]{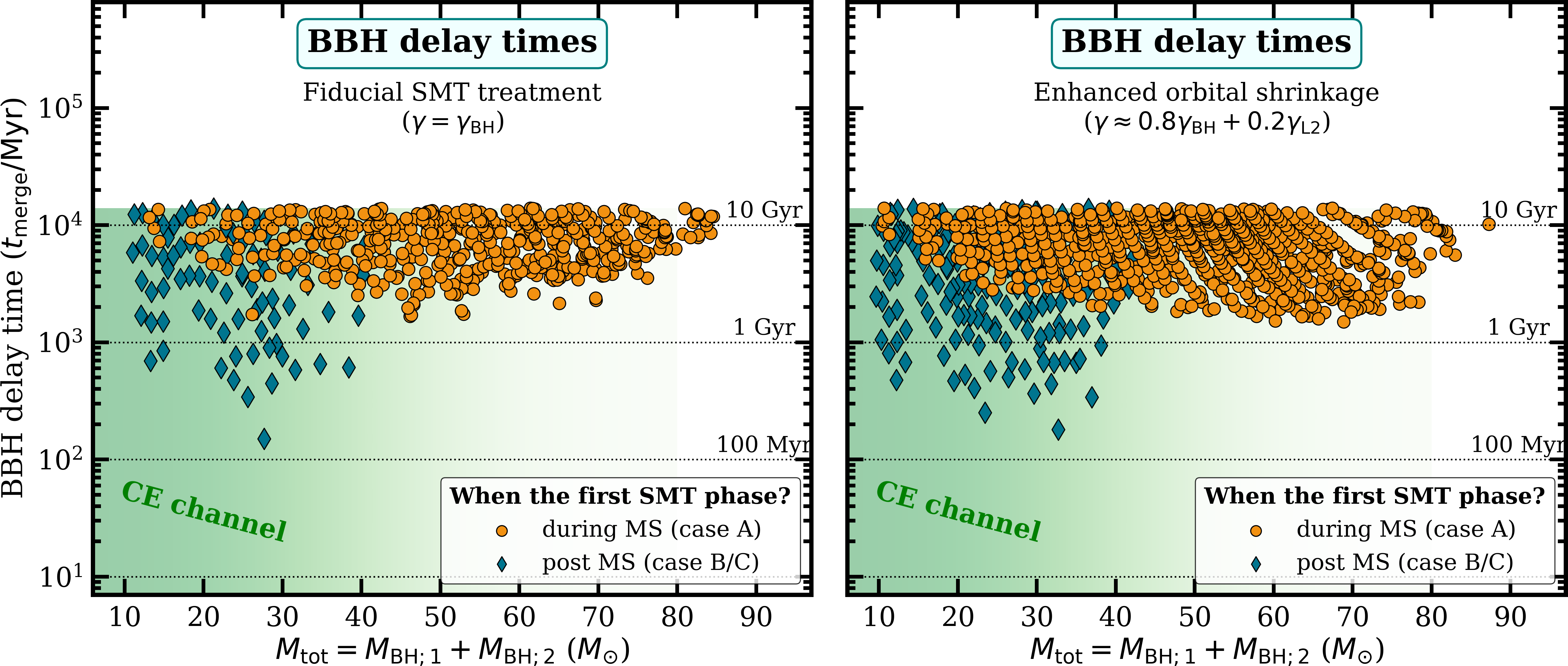}
    \caption{BBH mergers from the SMT channel have long delay times ($\gtrsim1$ Gyr) as a consequence of the fundamental separation limit, regardless of the uncertainties in binary orbit evolution (left vs right). Here, we plot the delay times of merging BBHs from SMT evolution ($t_{\rm merge} \lesssim 13.7$ Gyr) from all our binary MESA grids, as a function of the total BBH mass. 
    We distinguish between BBH mergers formed via SMT evolution from MS donors (case A mass transfer) and post-MS donors (case B or C). Notably, case A evolution dominates for massive BBH mergers ($\gtrsim40\Msun$). Two variations are considered: non-accreted mass ejected with the specific AM of the BH ($\gamma = \gamma_{\rm BH}$) and enhanced orbital shrinkage due to L2 outflows ($\gamma \approx 0.8 \gamma_{\rm BH} + 0.2 \gamma_{\rm L2}$). The BBH delay times are unaffected by the choice of $\gamma$. The green shading indicates the approximate range of delay times and masses where we expect the CE channel to contribute.
    The figure is based on MESA binary grids of BH+O-star systems with the firstly formed BH initial mass of $M_{\rm BH;1} = 4$, $7$, $10$, $13$, $16$, $20$, $24$, $28$, $32$, $36\Msun$, and $40\Msun$ (see Fig.~\ref{fig.grid_analyt_vs_MESA} for details and App.~\ref{sec.app_addFigs} for other masses).
    }
    \label{fig.bbh_delaytimes}
\end{figure*}

The MS donors are weakly affected by the change in semiconvective mixing (Fig.~\ref{fig.obywatel_gc}) or the $\gamma$ variation (Fig.~\ref{fig.triumvirate}). We find that most MS donors with $M \gtrsim 20\Msun$ and $R > 10\Rsun$ can lead to BBH mergers via SMT evolution (at $Z = 0.1\Zsun$ here), although with long delay times of a few Gyr. In fact, case A mass transfer may be the dominant mode of formation of BBH mergers in SMT evolution, especially above the total BBH mass of $40-50 \Msun$ (Fig.~\ref{fig.bbh_delaytimes}). While the BBH delay times from MS donors are generally $\gtrsim 2$ Gyr, the delay times from CHeB donors may be as short as $\sim 300$ Myr. That means that counter-intuitively, the most narrow BH+He-star orbits may be formed from initially the widest BH+O-star binaries with $M_{\rm O;Star} > M_{\rm BH}$ (Fig.~\ref{fig.slice}).\footnote{This is the reason for the trend in orbital separations of BH+He-star systems in Fig.~\ref{fig.fig3_dodgerblue}. }

We find that there are two types of MS donors that do not lead to BBH mergers (Fig.~\ref{fig.obywatel_gc}). The first are O-type stars in very close BH binaries that overflow their Roche lobes with $R_{\rm RLOF} \lesssim 10\Rsun$ (periods $\lesssim 2$ days). 
These donors generally engage into mass transfer when they are still early during their MS evolution. As a result, their He-core mass decreases with respect to a single star or a post-MS donor \citep{Schurmann2024}. This leads to the orbit re-expanding in the later half of the case A mass transfer due to the mass ratio reversal (see also the left panel of Fig.~\ref{fig.slice}), and BBH delay times significantly above $14$ Gyr. We note that this late re-expansion of the orbit can be reduced with enhanced orbital AM loss, e.g. due to L2 outflow (comparing left and right of Fig.~\ref{fig.triumvirate}).
While the regime of $\sim 1$ day orbits may not be the most suited for the SMT channel, it could be where the CHE channel for BBH mergers takes over \citep{deMink2016,Mandel2016,Marchant2016}. 

Finally, we find that very massive stars ($\gtrsim 80\Msun$) may not be suitable as donors for the SMT channel once they expand beyond $\sim50-100 \Rsun$. This comes as a surprise, given that these stars are characterized by the most extreme critical mass ratios of even up to 9 (top of Fig.~\ref{fig.triumvirate}). In principle, they should thus be the ideal candidates for enormous orbital shrinkage during the SMT phase and the formation of BBH mergers (as also found in our semi-analytical model). However, what detailed MESA models reveal is that the very massive stars do not transfer much mass before detaching from the SMT phase, never allowing the orbit to shrink significantly. This is because their envelopes are already He-enriched even without any mass transfer due to more extended MS convection and stronger mass loss. As a result, stars of $\gtrsim 80\Msun$ transfer only $\sim 20-30\%$ of their mass before contracting to $\lesssim 20 \Rsun$, which in wider orbits leads to an early detachment from mass transfer. 
This amount of transferred mass is far less than what is assumed is any rapid binary evolution model. 
After detachment, the partially-stripped He-enriched star is a hybrid between a normal single star and a fully stripped helium core, making it extremely difficult to model without detailed 1D stellar computations. This phenomena of the mass transfer stopping early was discovered by \citet{Sen2023} as a scenario to form reverse Algol systems, i.e. post-interaction binaries in which the mass-loser remains more massive than the mass gainer. Observationally, such stars would likely appear as H-rich WR stars of the WN type \citep{Pauli2022}. In the context of binary evolution scenarios, an early detachment from mass transfer would have significant consequences. It may indicate that binary interactions do not have as strong an effect on the properties and orbits of very massive stars in $\gtrsim 100$ day systems as they do on normal massive stars systems even up to $\sim1000$ days. Binary channels, such as the SMT pathways to BBH mergers, may become less effective in the regime of very massive stars. Without a better understanding of internal chemical mixing and stellar winds of very massive stars it may be impossible to arrive at robust predictions for GW sources from binary channels \citep[for a discussion of these challenges in the CE channel see][]{Romagnolo2025}.

\subsection{Consequences for the properties of BBH mergers}
\label{subsec.res_consequences_BBHmergers}

So far in Sec.~\ref{subsec.res_rapid_vs_detail}, we examined the formation of BBH mergers from BH+O-star MESA models with $M_{\rm BH;1} = 10\Msun$. We found BBH delay times of $\gtrsim 1$ Gyr, regardless of the degree of orbital shrinkage during SMT and the assumption on $\gamma$ (Sec.~\ref{subsec.res_there_is_a_limit}). We argued that the key factors determining BBH orbits in the SMT channel are instead the evolutionary stage and envelope structure of donor stars (Sec.~\ref{subsec.res_it_is_the_star_that_matters}). Here, we combine results from MESA grids of BH+O-star systems with other BH masses: $M_{\rm BH;1} = 4$, $7$, $10$, $13$, $16$, $20$, $24$, $28$, $32$, $36\Msun$, and $40\Msun$ (see Fig.~\ref{fig.appendix_4_7_10} for a collection of grid figures, all at $Z = 0.1\Zsun$ metallicity). 
For each $M_{\rm BH;1}$ case, we simulated a range of periods and mass ratios that is large enough to capture the entire parameter space for the formation of BBH mergers (notably, the range changes with mass, Fig.~\ref{fig.appendix_4_7_10}). Two variations were explored: the fiducial SMT treatment with $\gamma = \gamma_{\rm BH}$, and variation with enhanced orbital shrinkage due to AM loss through L2 outflows ($\gamma \approx 0.8 \gamma_{\rm BH} + 0.2 \gamma_{\rm L2}$). 
As we do not model the prior distribution of BH + O-type systems formed in Nature, we do not make predictions for the formation rate or the mass distribution of the observable population of BBH sources (see however the discussion in Sec.~\ref{sec.disc_gw_implications}).
Instead, in the following Fig.~\ref{fig.bbh_delaytimes} and Fig.~\ref{fig.bbh_BRW}, we demonstrate what is the entire possible range of BBH masses, delay times, and mass ratios that may be realized in the SMT channel given our assumption of direct collapse with no natal kick (in Sec.~\ref{sec.disc_kicks} we discuss how BH kicks may affect our results). 

In Fig.~\ref{fig.bbh_delaytimes}, we show the BBH delay times as a function of the total BBH mass for all the BBH mergers formed across our MESA binary grids. 
In two panels we compare both $\gamma$ variations, confirming that the enhanced orbital shrinkage from L2 outflows does not lead to shorter delay times across all the BBH masses. We distinguish between BBH mergers formed via SMT evolution from MS donors (case A mass transfer) and post-MS donors (case B or C). Green shading indicates where we expect the CE channel to contribute, assuming that it falls off at the high-mass end as indicated by the empirical lack of red supergiants with masses $\gtrsim 40\Msun$ \citep{Davies2018}, see also \citet{Klencki2021}.
For the SMT channel, we find that MS donors always leads to long BBH delay times of $\gtrsim 2$ Gyr, whereas post-MS donors (specifically CHeB giants) may lead to tighter BBH systems and shorter delay times of a few hundred Myr. This is in line with our investigation of different donor types and structures (Sec.~\ref{subsec.res_it_is_the_star_that_matters}, see also discussion in Sec.~\ref{subsec.disc_why_the_limit_exists}). In addition, we find that BBH mergers with the total mass above $\sim 40\Msun$ form predominantly from case A mass transfer. This is because of the large radii of MS stars above the mass of $ \sim50-60\Msun$ in our models, such that the case A interaction occurs even in wider orbits of hundreds of $\Rsun$ (Fig.~\ref{fig.obywatel_gc}).
While the radii of these massive MS stars are likely overly inflated in 1D stellar models \citep{Sanyal2015,Sanyal2017,Klencki2020,Agrawal2020,Romagnolo2023}, we find that very massive stars in wider binaries (i.e. typical case B/C range) are not efficient in forming BBH mergers via the SMT channel due to their He-rich envelopes causing an early detachment (Fig.~\ref{fig.obywatel_gc}). Our results suggest that case A evolution may be the dominant formation channel of massive BBH mergers via SMT evolution. This poses a challenge for the SMT treatment in rapid binary evolution codes to accurately predict the cores and final masses of case A donor stars \citep[][]{Belczynski2022,RomeroShaw2023,Schurmann2024,Shikauchi2024}.

\begin{figure*}
    \includegraphics[width=\textwidth]{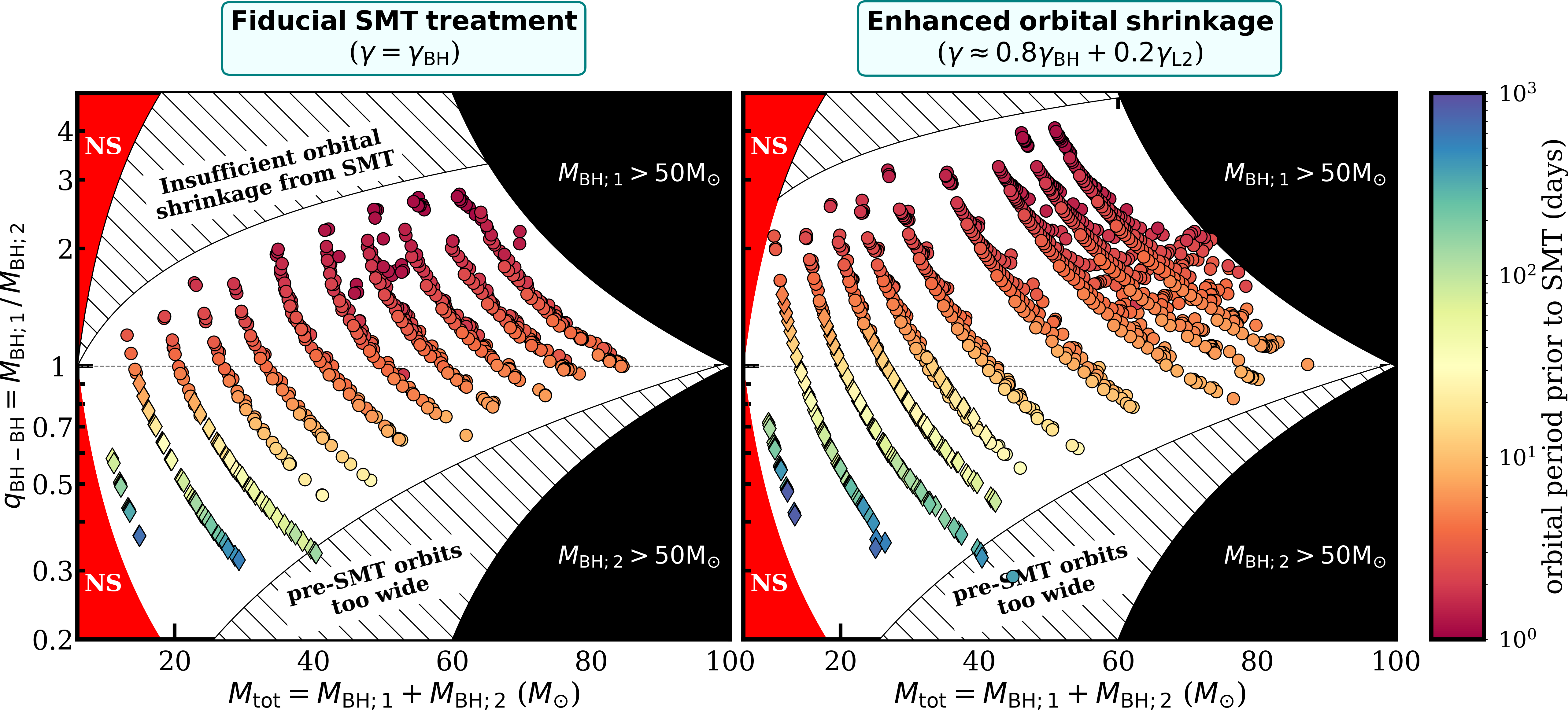}
    \caption{The mass ratios of BBH mergers from SMT evolution are confined to the range from $\sim0.3$ to $\sim3$, determined by the physics of stable binary interactions. We plot the BBH mass ratios $q_{\rm BH-BH}$ as a function of the BBH mass from our binary MESA grids, where $q_{\rm BH-BH}$ is defined as $M_{\rm BH;1} / M_{\rm BH;2}$ with $M_{\rm BH;1}$ ($M_{\rm BH;2}$) being the mass of the first (second) born BH. Hatched regions indicate the combinations of BBH masses and mass ratios that cannot be produced via SMT evolution (see text). This results in a positive trend between $q_{\rm BH-BH}$ and $M_{\rm tot}$. Circles indicate models in which the SMT interaction was initiated by a MS donor (case A) and diamonds mark SMT evolution from a post-MS donor (case B or C). The color indicates the orbital period at the BH+O-star stage (i.e. before the SMT). Based on binary grids of BH+O-star systems with $M_{\rm BH;1} = 4$, $7$, $10$, $13$, $16$, $20$, $24$, $28$, $32$, $36\Msun$, and $40\Msun$.}
    \label{fig.bbh_BRW}
\end{figure*}

In Fig.~\ref{fig.bbh_BRW}, we examine the mass ratios of BBH mergers as a function of their total mass. We define the mass ratio $q_{\rm BH-BH}$ as $M_{\rm BH;1} / M_{\rm BH;2}$, meaning that $q_{\rm BH-BH} > 1$ are systems in which the firstly formed BH is the more massive, and vice versa. Similarly to Fig.~\ref{fig.bbh_delaytimes}, we include BBH mergers formed across all our BH+O-star grids. They are arrangement along several discrete curves, each corresponding to a different $M_{\rm BH;1}$, ranging from $4$ to $40\Msun$.
% Their color indicates the orbital period at the BH+O-star stage, i.e. before the SMT phase. 
In black, we mark BBH systems in which one of the BHs would exceed the pair-instability limit (\citealt{Farmer2019,Renzo2022}, although see \citealt{Farag2022}), while red is the regime of BH + NS mergers assuming an upper NS mass limit at $3\Msun$.
Hatched regions indicate the combinations of BBH masses and mass ratios that are not produced by SMT evolution according to our models. We find that the exclusion of high $q_{\rm BH-BH}$ systems (i.e. the secondary-formed BH being notably less massive) is due to insufficient orbital shrinkage during SMT. For BBH mergers to form with such mass ratios, the companion stars at the BH+O-star stage could not be too massive with respect to the BH, which prevents the orbit from shrinking during SMT. This limitation is partly alleviated if enhanced orbital shrinkage is assumed (the right panel), though even in this case none of the BBH mergers exceed $q_{\rm BH-BH} \approx 3$. More extreme BBH mass ratios could potentially be obtained if a substantial amount of mass is lost during the formation of the BH (i.e. non-direct collapse), although the resulting Blaauw kick would further increase the already long delay times that we find (see Sec.~\ref{sec.disc_kicks}).
On the other end of the spectrum, BBH mergers with very low $q_{\rm BH-BH}$ would require BH+O-star systems in which the stellar companion is substantially more massive than the BH, leading to significant orbital shrinkage during the SMT phase and a wide pre-SMT orbit. We find that such BBHs do not form in the SMT channel, particularly at high masses, because interactions with very massive donors ($\gtrsim 75\Msun$) in wide orbits do not lead to close post-SMT orbits (see Fig.~\ref{fig.obywatel_gc} and donors with He-enriched envelopes).

Fig.~\ref{fig.bbh_BRW} reveals a positive trend between $q_{\rm BH-BH}$ and $M_{\rm tot}$. However, it is unclear to what extend this effect might be realized in the measured population of BBH merges. Observationally, it is impossible to infer which BH was formed first, meaning that the $q_{\rm BH-BH} > 1$ values would instead be measured as $1.0 / q_{\rm BH-BH}$, negating the existence of any trends.

\section{Discussion}
\label{sec.discussion}

\subsection{Why is there a limit on post-SMT separations?}
\label{subsec.disc_why_the_limit_exists}

In this section, we discuss the following questions: 
(a) Why is there a limit on how close the post-SMT orbital separation can be and what determines the $\sim8\Rsun$ value? 
(b) Why does the limit remain unchanged even if enhanced orbital shrinkage due to strong AM loss through L2 is assumed?

\begin{figure}
    \includegraphics[width=\columnwidth]{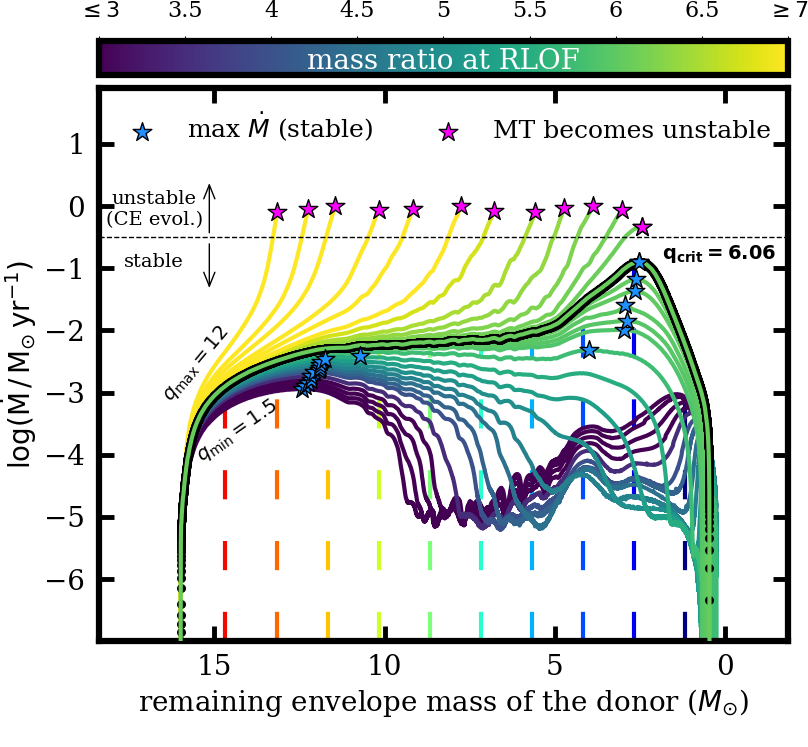}
    \vspace{0.1cm}
    \includegraphics[width=\columnwidth]{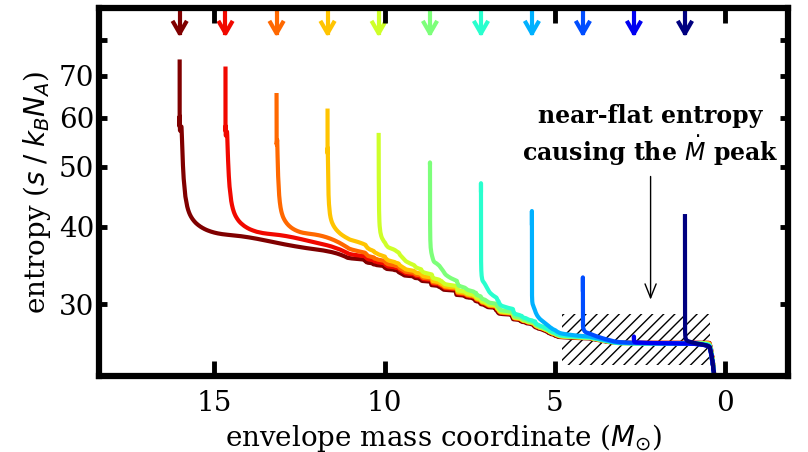}
    \includegraphics[width=\columnwidth]{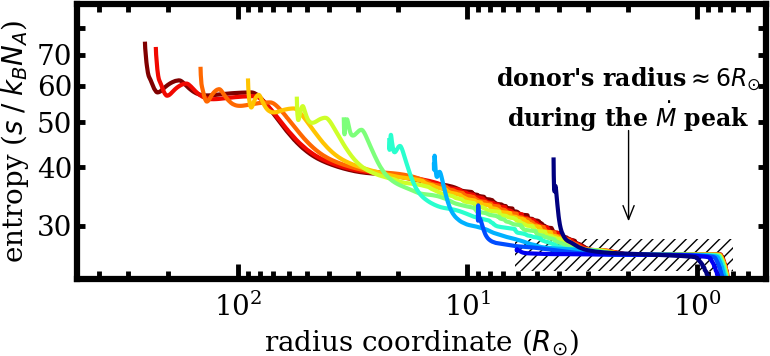}
    \caption{The origin of the separation limit lies in delayed dynamical instability. \textbf{Top:} mass transfer rate ($\dot{M}$) in systems with a $30\Msun$ donor and a varying mass ratio at RLOF, shown as a function of the decreasing envelope mass. The donor is the same for each mass ratio at the onset of RLOF: $\sim$250$\Rsun$ core-He burning giant with a $\sim$17$\Msun$ envelope and $\sim$13$\Msun$ core.
    Models reaching ${\rm log} (\dot{M} / {\rm M_{\odot} yr^{-1}}) > 0.5$ become unstable (Sec.~\ref{subsec.method_mesa}). The critical mass ratio for stability is $q_{\rm crit} = 6.06$ (marked in bold). The threshold between stable and unstable models is determined by the $\dot{M}$ peak at $M_{\rm env} \approx 2.5 \Msun$.
    \textbf{Middle:} internal entropy profiles of the donor in the $q =q_{\rm crit}$ model, ranging from the onset of RLOF (dark red, left) to the end of mass transfer (dark blue, right). The $\dot{M}$ peak seen in the top panel is caused by stripping off the layers where the entropy profile is nearly flat. \textbf{Bottom:} same entropy profiles but as a function of the radius coordinate. The flat entropy layers are within the inner $3\Rsun$ of the donor ($6\Rsun$ during the $\dot{M}$ peak), causing instability in models in which the mass transfer would shrink the orbit below $a \lesssim 8\Rsun$.
    }
    \label{fig.disc_point_prenvelopeofiles}
\end{figure}

In Fig.~\ref{fig.disc_point_prenvelopeofiles} we examine mass transfer evolution from a $30\Msun$ donor interacting in systems with BH companions of different masses. The mass ratio at RLOF $q_{\rm RLOF} = M_{\rm donor} / M_{\rm BH}$ varies from $q_{\rm min} = 1.5$ to $q_{\rm max} = 12$. The donor is the same for each $q_{\rm RLOF}$: a core-He burning giant of $252\Rsun$ at $0.1\Zsun$ metallicity with a $\sim17\Msun$ envelope and a $\sim13\Msun$ core.\footnote{These models are a small subset of the grid with a varying $q_{\rm RLOF}$ that was computed to determine the critical mass ratio $q_{\rm crit}$ for the donors of different masses and radii in Fig.~\ref{fig.triumvirate}.} We chose this particular donor as it leads to the smallest post-SMT orbit of $\sim 8\Rsun$. The top panel in Fig.~\ref{fig.disc_point_prenvelopeofiles} shows the mass transfer rate evolution as a function of the decreasing envelope mass. Models reaching ${\rm log} (\dot{M} / {\rm M_{\odot} yr^{-1}}) > 0.5$ are assumed to become unstable. The model with $q_{\rm RLOF} = q_{\rm crit} = 6.06$ (shown in bold) is the critical one, i.e. at the threshold between stable and unstable mass transfer evolution, leading to the smallest post-SMT orbital separation. In the remaining two panels of Fig.~\ref{fig.disc_point_prenvelopeofiles}, we show the internal entropy profiles of the donor star from the critical model, plotted every $1.5\Msun$ from the onset of RLOF (dark red, left) to near the end of the mass transfer (dark blue, right).  

We analyze how the mass transfer rate $\dot{M}$ evolves across models of different mass ratios. Systems with $q_{\rm RLOF} \leq 6.06$ remain stable. For most stable mass ratios ($q_{\rm RLOF} < 5.5$), the $\dot{M}$ peaks once about $30\%$ of the envelope has been transferred (here at the envelope mass $\sim12\Msun$). In the remainder of the mass transfer phase, the rate is smaller and may even drop to the nuclear-timescale $\dot{M} \approx 10^{-5} \Msunyr$. This is consistent with prior models of stable mass transfer evolution \citep[e.g.][]{Podsiadlowski2002,Rappaport2005,Pavlovskii2017,Klencki2022}. However, as the mass ratio of the model increases ($q_{\rm RLOF} > 5.5$), we begin to observe a delayed $\dot{M}$ peak appear at late stages of the interaction, when the final $\sim5\Msun$ of the envelope are being transferred. The delayed $\dot{M}$ peak rises in magnitude the higher the mass ratio and eventually triggers instability for $q_{\rm crit} = 6.08$. 
Notably, among models with $q_{\rm RLOF}$ near the $q_{\rm crit}$ value, the peak mass transfer rate varies strongly as a function of $q_{\rm RLOF}$. For example, comparison models with $q_{\rm RLOF} = 5.98$, $6.08$ (critical), and $6.12$, their maximum reached $\dot{M}$ values are $\approx 0.025$, $0.34$, and $1.1 \Msunyr$, respectively. This indicates that our choice to set the critical rate $\dot{M}_{\rm crit}$ at $10^{-0.5} \Msunyr$  has a small effect on $q_{\rm crit}$, at least for post-MS donors. For the MS donors the uncertainty on $q_{\rm crit}$ due to instability criteria is larger.

Fig.~\ref{fig.disc_point_prenvelopeofiles} reveals that is the delayed $\dot{M}$ peak at $M_{\rm env} \approx 2.5\Msun$ that determines the transition from stable to unstable MT evolution and the critical mass ratio. By examining the donor profiles in the middle panel, it becomes clear that this $\dot{M}$ peak is caused by the bottom envelope layers in which the entropy profile is nearly flat. This is a well-known effect: the flatter the entropy profile, the more the donor star expands in response to mass loss, leading to higher mass transfer rates \citep[for a perfectly constant entropy, the donor would react as ${\rm dln}R/{\rm dln}M = -1/3$,][]{PaczynskiZiolkowskiZytkow1969,PaczynskiSienkiewicz1972,Webbink1985}. Mass transfer instability caused by a flattening  entropy profile in deep stellar interiors has been identified as delayed dynamical instability already by \citet{Hjellming1987}.

\begin{figure}
    \includegraphics[width=\columnwidth]{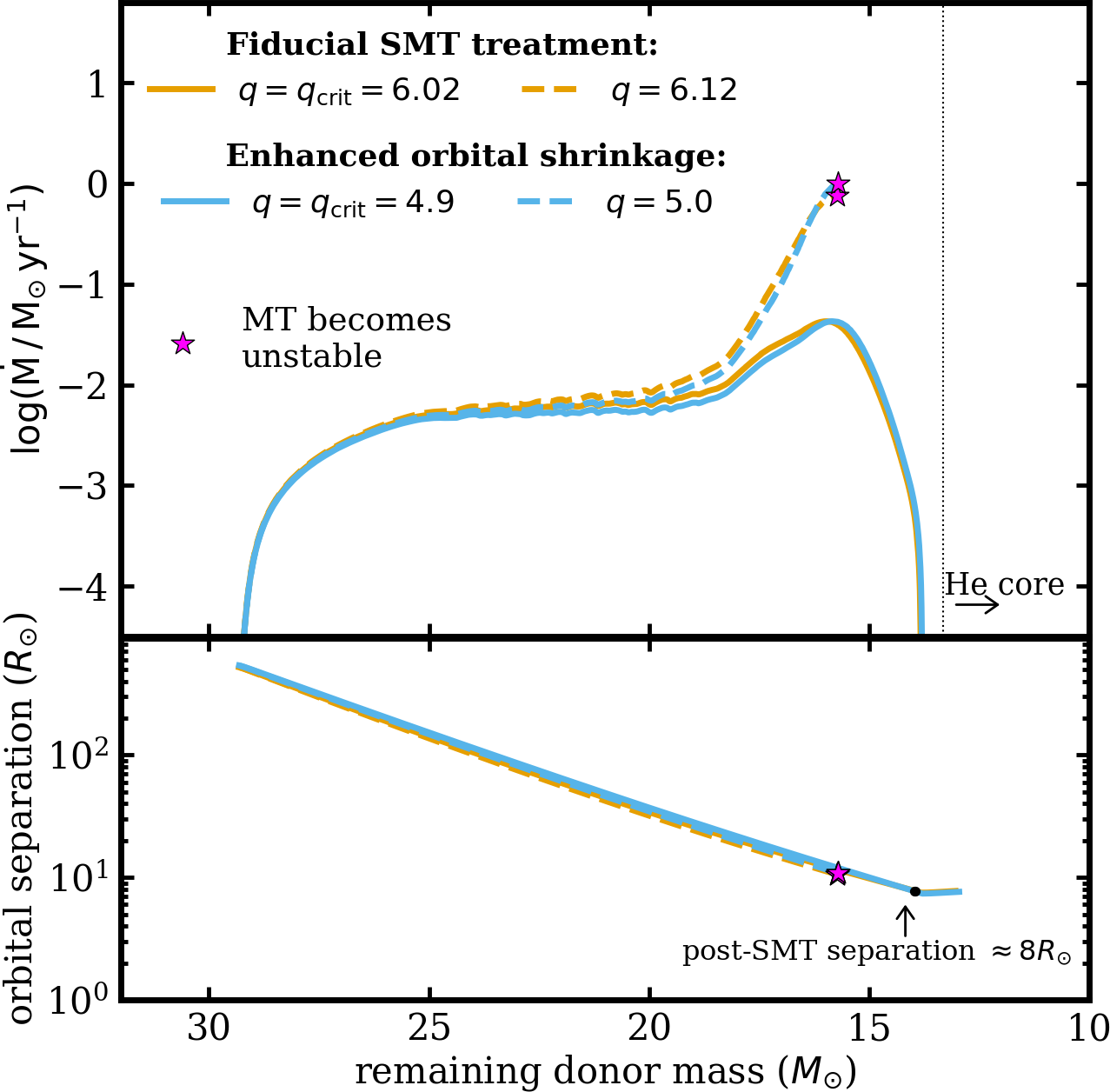}
    \caption{The path to delayed dynamical instability and the resulting minimum post-SMT separation do not depend on the assumption on orbital shrinkage ($\gamma$ value). Here, we plot the evolution of the mass transfer rate (top) and orbital separation (bottom) in binaries interacting at the critical mass ratio for the fiducial SMT treatment ($\gamma = \gamma_{\rm BH}$) and assuming enhanced shrinkage due to L2 outflows ($\gamma \approx 0.8 \gamma_{\rm BH} + 0.2 \gamma_{\rm L2}$). The donor is the same $30\Msun$ CHeB giant as in Fig.~\ref{fig.disc_point_prenvelopeofiles}, the companion is a BH accreting at the Eddington limit, with its mass determined by the $q_{\rm crit}$ value: $\approx5\Msun$ in the fiducial model, $\approx 6\Msun$ in the case with enhanced shrinkage.     
    Regardless of the assumed $\gamma$, the critical $\dot{M}$ peak that determines stability occurs at the same moment of the interaction. As a result, the mass and the radius of the donor at this critical point do not depend on $\gamma$, leading to the same final orbital separation. }
    \label{fig.disc_MdotEvol_FidvsL2}
\end{figure}

The near-flat entropy layers are present in the inner $\sim5\Msun$ of the donor's envelope already at the moment of RLOF (dark red profile), enclosed within just the inner $\sim3\Rsun$ of the $252\Rsun$ giant (the bottom panel). These layers remained nearly unchanged through most of the interaction and hidden deep inside the Roche lobe: as the mass transfer progresses (profiles from left to right), the entropy adjusts in the outer layers of the donor such that the near-surface entropy peak, characteristic of radiative envelopes, is rebuilt.\footnote{As pointed out already by \citet{PaczynskiSienkiewicz1972} and more recently discussed by \citet{Podsiadlowski2002,Woods2011,Pavlovskii2015}, the entropy adjustment in the outermost layers of the donor plays a meaningful role in how the donor responses to mass loss, decreasing $\dot{M}$ and stabilizing the mass transfer, which is neglected in the classical adiabatic results such as \citet{Hjellming1987}.} However, as the donor becomes progressively more stripped and smaller in size, the flat entropy layers come closer to the surface. Eventually, if the orbit has shrunk enough and the flat entropy layers begin to occupy a substantial volume of the Roche lobe, they cause the delayed $\dot{M}$ rise. Its peak value $\dot{M} \approx 0.1 \Msunyr$ at $M_{\rm env} \approx 2.5\Msun$ corresponds to the penultimate profile in the middle panel (indicated by the arrow). For such a high transfer rate, the donor loses mass on a timescale that is shorter than the thermal-timescale of even its outermost envelope, such that the response becomes nearly adiabatic and the subsurface entropy peak is no longer rebuilt \citep{Temmink2023}. At that point, the size of the donor is about $6\Rsun$ (orbital separation $\approx 12\Rsun$) and its entropy profile is nearly flat all the way to the surface (bottom panel). This is the critical bottleneck that determines how tight the post-SMT orbit can be. To avoid instability, the orbit cannot shrink significantly while the flat entropy layers are being transferred. For the donor in Fig.~\ref{fig.disc_point_prenvelopeofiles}, the Roche-lobe exponent must remain $\zeta_{\rm RL} \lesssim 4$ or $dR_{\rm RL}/dM_{\rm don} \lesssim 1.5$ during the delayed $\dot{M}$ peak. As a result, from that point on, the orbital size only slightly decreases from $\sim12\Rsun$ down to $\sim8\Rsun$ at the end of the interaction. Importantly, the properties of the donor (mass, radius) at the critical point of max $\dot{M}$, and consequently the orbital separation, are the same no matter what is the assumption on the AM budget of the SMT phase ($\gamma$): see Fig.~\ref{fig.disc_MdotEvol_FidvsL2}. This is why throughout the paper we find that the minimum post-SMT separations and the BBH orbits are independent of $\gamma$. The same considerations imply that they are also independent of how much matter is accreted ($\beta$).\footnote{The only hypothetical way to have these uncertainties play a role is if $\gamma$ was not constant but would increase only after the flat entropy layers have been stripped, in which case the transfer of the remaining envelope could lead to stronger orbital shrinkage at late SMT stages. This would require the specific AM loss increasing with the decreasing mass transfer rate.} Our results are qualitatively similar to those of \citet{Ge2015} who describe that the critical mass ratios for radiative donors are set by a delayed dynamical instability triggered in layers where the entropy is nearly constant.

The discussion of Fig.~\ref{fig.disc_point_prenvelopeofiles} suggests that it is the presence of flat entropy layers at the bottom of the radiative envelope of CHeB giants that prevents the formation of $\sim2-8 \Rsun$ orbits via SMT evolution, which consequently lead to long BBH delay times. A question arises how robust is the presence of such layers in the the envelope and whether they could be avoided. In Sec.~\ref{sec.disc_why_donors_differ}, we compare the internal structures of CHeB giants with other main donor types in our grids, finding that layers of flat entropy are always present and that the case explored in Fig.~\ref{fig.disc_point_prenvelopeofiles} is the most favorable for the formation of short-period orbits, as well as discuss whether hypothetical 'ultra-stable' donor stars could exist in Nature.

\subsection{Why are some donor types favorable for the SMT channel? The role of chemical mixing for stability}
\label{sec.disc_why_donors_differ}

In Sec.~\ref{subsec.res_it_is_the_star_that_matters}, we found that the minimum size of the post-SMT orbit depends on the type of donor star, with smallest separations achievable from CHeB giants ($\sim8-15\Rsun$) and MS donors ($\sim15-30\Rsun$), and wider orbits from HG giants and convective-envelope stars. Here, we discuss whether those findings can be understood as a result of differences in the internal structure of different donors. 

\begin{figure*}
    \includegraphics[width=\textwidth]
    {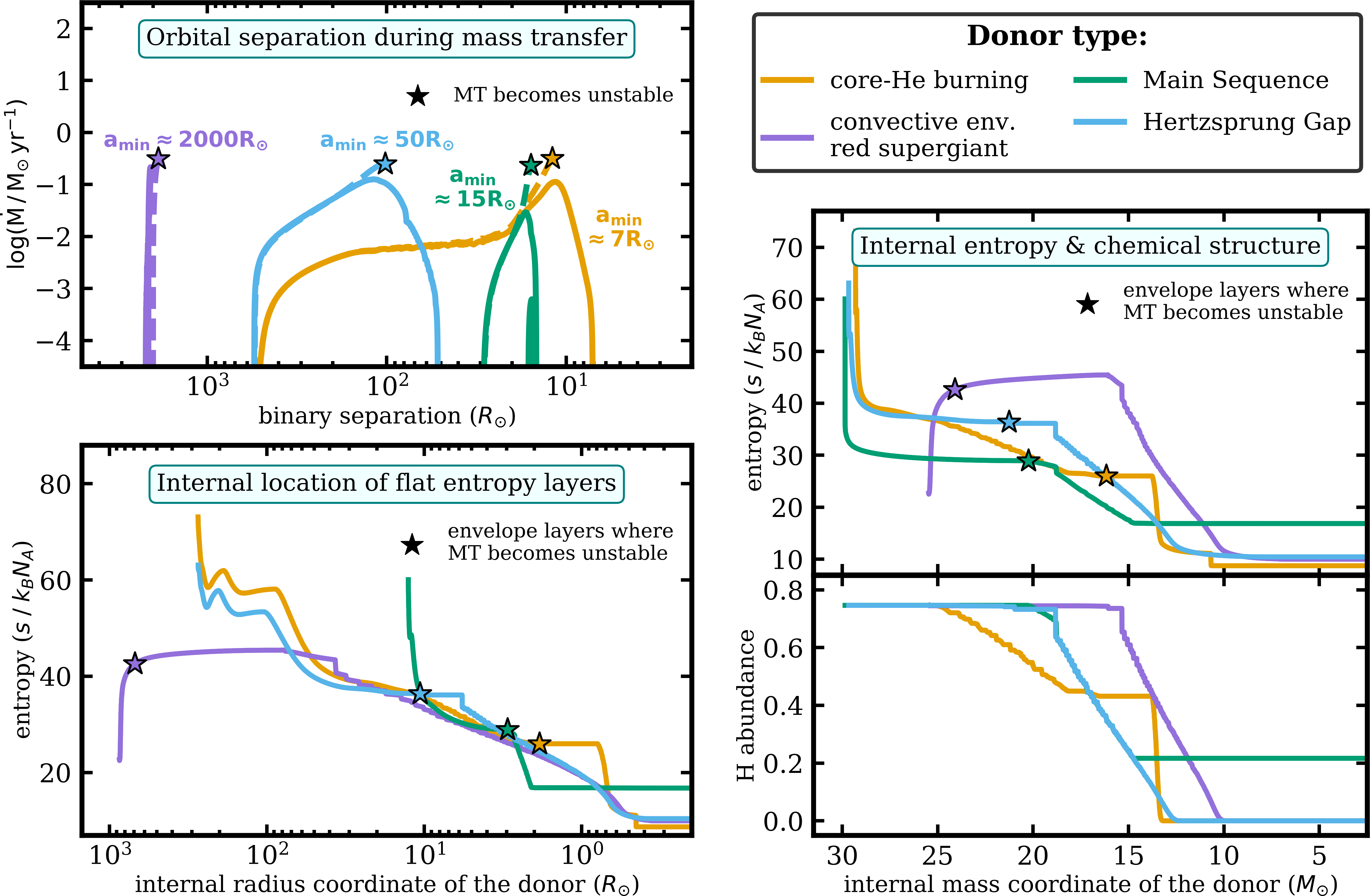}
    \caption{The minimum orbital separation from SMT evolution is determined by the stellar structure of the donor. \textbf{Top left:} evolution of the mass transfer rate ($\dot{M}$) as a function of the decreasing binary separation in interacting systems with a BH accretor and a $30\Msun$ donor of different donor types. The CHeB donor is the same $\sim 250\Rsun$ giant as in Fig.~\ref{fig.disc_point_prenvelopeofiles}, whereas its HG variant comes from the grid with low semiconvection. The solid lines are models at $q = q_{\rm crit}$, dashed lines are at $q$ slightly larger than $q_{\rm crit}$, such that the MT becomes unstable (star symbols). \textbf{Right:} Internal entropy and H abundance profiles of different donors from the onset of RLOF, plotted as a function of the internal mass coordinate $m$. The He-core in post-MS donors is at $m \lesssim 13\Msun$. The layers that cause the $\dot{M}$ peak and dynamical instability (star symbols) are where the entropy profile is nearly flat. \textbf{Bottom left:} Same entropy profiles but as a function of the radial coordinate. The difference in the location of the constant entropy regions in the envelopes of different donors is primarily responsible for different minimum orbital separations $a_{\rm min}$ that may be reached via SMT.}
    \label{fig.disc_different_donors}
\end{figure*}

In the top left panel of Fig.~\ref{fig.disc_different_donors}, we examine the evolution of the mass transfer rate $\dot{M}$ as a function of the decreasing binary separation for four main donor types, comparing the evolution at $q = q_{\rm crit}$ (solid lines, stable MT) and at $q$ slightly larger than $q_{\rm crit}$ (dashed lines, unstable MT). The donor mass is $30\Msun$, with a core mass of $\sim 13\Msun$ (for post-MS donors), or $26\Msun$ with the core of $\sim10\Msun$ for the convective-envelope donor (to avoid lines in other panels overlapping). The CHeB donor is the same $\sim250\Rsun$ giant as in Fig.~\ref{fig.disc_point_prenvelopeofiles}, while the HG donor is taken from the grid variation with low semiconvection ($\alpha_{\rm SC} = 0.01$, see Fig.~\ref{fig.obywatel_gc}). For each donor, we label the final orbital separation from SMT evolution at $q = q_{\rm crit}$. The figure illustrates that, depending on the donor type, the critical mass transfer rate peak that may lead to instability (star symbols) occurs at different stages of the interaction. For example, for convective donors the instability is triggered shortly after the onset of RLOF, whereas for CHeB donors it is substantially delayed, as the $\dot{M}$ peaks when the separation as well as the donor are much smaller than at the start of the mass transfer.

It turns out that for each donor type, the point when the mass transfer becomes unstable is related to where in the envelope of the donor the entropy profile becomes nearly flat. 
We see this in the other panels of Fig.~\ref{fig.disc_different_donors}, which show the entropy profiles of different donors at the onset of RLOF, plotted as a function of the mass coordinate (top right) or the radius coordinate (bottom left). The star symbols indicate precisely which layers are being transferred at the moment when the MT becomes unstable. In the bottom right panel we show the internal H abundance profiles for comparison, as it is the chemical composition structure that determines the entropy structure of a radiative envelope.

For the MS donor in Fig.~\ref{fig.disc_different_donors}, the instability occurs at $m \approx20 \Msun$, at the bottom of the outer part of the star with the pristine H abundance. This is where the entropy profile is the flattest, besides the inner convective core ($m < 15\Msun$). The region with $15 < m/\Msun < 20$ is characterized by an entropy slope, which stabilizes the mass transfer and leads to lower $\dot{M}$ rates compared to the peak at $m \approx 20\Msun$. This entropy slope is a direct consequence of the H abundance slope that was created by the  convective core decreasing in size during the prior MS evolution. The gradient in the H abundance introduces a gradient in the mean molecular weight, which in turn causes a gradient in the specific entropy: the most relevant quantity for how the donor star responses to mass loss. In that way, the abundance gradients in stars have a stabilizing effect on the mass transfer. The entropy and the abundance profiles of the HG donor are similar to the MS donor, with the same H abundance gradient present on top of the now fully H-depleted helium core. As a result, the instability occurs at the similar layer of $m \approx 21\Msun$. The striking difference, however, is in the radial coordinate at which the flat entropy layers are located (bottom left in Fig.~\ref{fig.disc_different_donors}). The envelopes of HG stars have undergone a phase of substantial expansion following the end of the MS. As a result, the near-constant entropy region extends out to the radius of $\approx 30-40\Rsun$, compared to $\approx 3-4\Rsun$ in the MS donor. For the HG donor, the $\dot{M}$ peaks when the binary separation is still quite large ($\sim 100\Rsun$). To avoid instability, the orbit cannot shrink too quickly at this point, which limits the final separation at a few tens of $\Rsun$. This is also where the difference between HG and CHeB donors becomes apparent. First, the H abundance profile of the CHeB giant has been more significantly altered compared to the MS star due to more extensive mixing via (semi)convection in the envelope layers above the H-burning shell \citep{Schootemeijer2019,Klencki2020,Kaiser2020}. This has extended the He-enriched layers up to $m \approx 25\Msun$, inducing a mean molecular weight gradient and consequently an entropy gradient that decreases the $\dot{M}$ rate compared to the HG donor case and stabilizes the mass transfer. On the other hand, the CHeB giant developed an intermediate convective zone at the bottom of the envelope, which flattened the H abundance and the entropy profiles in layers $14 < m/\Msun < 18$. These are the envelope layers responsible for the delayed $\dot{M}$ peak observed in the top left panel (and in Fig.~\ref{fig.disc_point_prenvelopeofiles}) that determines the critical mass ratio and the minimum post-SMT separation. Notably, they are located far closer to the core than in the HG giant case, at radial coordinates of $\sim 3\Rsun$, resulting in the peak of $\dot{M}$ occurring when the binary separation is already small ($\sim 10-15\Rsun$). This is why CHeB donors may lead to far tighter orbits via SMT evolution compared to HG donors. In the case of MS donors, the $\dot{M}$ peak also occurs when the orbit is quite small. However, what follows is the mass ratio reversal and orbital expansion, which is why MS donor lead to somewhat wider orbits than CHeB donors. 

The above analysis reveals an intriguing effect: chemical composition gradients induce mean molecular weight gradient, which in turn introduce specific entropy gradients that have a stabilizing effect on the mass transfer. This means that a hypothetical ultra-stable type of a donor star would be a radiative-envelope post-MS giant with no intermediate convective zone and with a He/H composition gradient extending throughout most of its envelope.

\subsection{Can GW astronomy teach us about stellar interiors and solve the blue-supergiant problem?}
\label{sec.disc_GW_stellar_interiors}

Fig.~\ref{fig.obywatel_gc} suggests that different assumptions on the efficiency of internal mixing in stars may have a dramatic impact on the formation of BBH mergers via SMT evolution. This is because the ability of SMT to form a close-orbit system is fundamentally related to the internal structure of the donor star (Sec.~\ref{subsec.disc_why_the_limit_exists}) and is robust against uncertainties in mass transfer (AM loss, accretion efficiency). 
This leads to a question whether the population of GW sources can teach us about stellar interiors. 
Indeed, while any prediction from stellar evolution (e.g. tracks, isochrones, metal yields) is always subject to uncertainties in the stellar physics (e.g. internal mixing and transport, nuclear reaction rates, mass loss), the effect illustrated in Fig.~\ref{fig.obywatel_gc} is exceptionally significant. Assuming low efficiency of semiconvection (the right panel), makes half the parameter space disappear (all post-MS donors), excludes most of $20-40\Msun$ stellar progenitors, and likely shifts the predicted BH mass distribution to higher masses by a considerable amount. 

Interestingly, the two variations of stellar evolution in Fig.~\ref{fig.obywatel_gc} are not arbitrary but are closely related to a long-standing puzzle of the nature of blue supergiants \citep{Langer1985,Langer1989,Stothers1992,Schootemeijer2019,Klencki2020,Bellinger2024}. 
For stellar models in the left panel, most massive stars spend a substantial fraction of their core-He burning lifetime as blue supergiants, which could explain the large number of blue supergiants found in spectroscopic surveys \citep{HumphreysMcElroy1984,Casares2014,Castro2018,deBurgos2023,Patrick2025}. We find that SMT evolution from such core-He burning blue supergiants may lead to BBH mergers. 
For stellar models in the right panel,
 massive stars tend to rapidly expand at the end of MS and spend most of their core-He burning as red supergiants, in which case the observed population of blue supergiants would be a manifestation of a more unique evolutionary scenario, possibly of stellar mergers or past mass accretors \citep{Vanbeveren2013,Justham2014,Bellinger2024,Menon2024,Schneider2024}. We find that SMT evolution from such rapidly expanding donors does not lead to BBH mergers. This suggests that in the future, a well characterized population of BBH mergers could help resolve the blue supergiants problem. 
\footnote{While we chose to vary semiconvection in Fig.~\ref{fig.obywatel_gc}, various mixing processes simultaneously play a role in establishing the chemical structure of the radiative envelope, which in turn determines the transition from blue to red supergiants \citep[e.g.][]{Georgy2013,Kaiser2020,Farrell2022} as well as the minimum separation from SMT evolution (Sec.~\ref{sec.disc_why_donors_differ})}.

While this is an exciting outlook, it should be noted that our study if just the first step. An important caveat is that, at the moment, we are dealing with single star models rather than models of past mass-gainer stars, in which the chemical abundance and entropy profiles will be different \citep{Renzo2021,Renzo2023,Schneider2024}.

\subsection{Implications for BBH mergers: delay times, masses, mass ratios, spins \& metallicity}
\label{sec.disc_gw_implications}

Here, we discuss the main implications of our findings for the properties of BBH mergers from the SMT channel.

\textbf{Delay times.} We find that because of the fundamental separation limit from SMT evolution, the BBH systems formed have preferentially long delay times ($\gtrsim 2$Gyr), independently of how much AM is lost during SMT (Fig.~\ref{fig.bbh_delaytimes}). This means that the contribution from the SMT channel is expected to fall of with redshift more quickly than in channels with shorter delay times, especially at z > 2-3 (lookback time $\sim 10$ Gyr). We find that shorter delay times of a few hundred Myr are possible for BBH mergers with masses $\lesssim 40 \Msun$. This could be the regime where the CE channel is the strongest due to the prevalence of red supergiants \citep{Klencki2021}. If the CE channel can produce shorter delay times of $\sim10-100$ Myr than there may be an overall trend in binary channels: the more massive the BBH, the longer the delay time.

\textbf{BBH masses.} We find that the SMT channel can operate in the entire explored range of BH masses from $\sim4\Msun$ up to the pair-instability mass gap at $\sim45\Msun$ (BBH total masses from $\sim10\Msun$ to $\sim90\Msun$, Fig.~\ref{fig.appendix_4_7_10}). The size of the parameter space is comparable for all BH masses above $\gtrsim 10\Msun$. This means that assuming a flat prior on mass ratios and log periods of BH+O-star systems, our models would likely favor the low-mass BH end (due to the initial-mass function), rather than the $\gtrsim20\Msun$ range as found in some population models \citep{Olejak2022,vanSon2022a}. On the other hand, we show that the range of donor masses that facilitate the SMT channel depends on the assumed stellar physics of internal mixing (Fig.~\ref{fig.obywatel_gc}): different assumptions on the structure of stars may affect the mass distribution considerably. Regardless, we find that case A mass transfer at the BH+O-star stage dominates the formation of BBH mergers with masses $\gtrsim40\Msun$. It is also worth pointing out that the parameter space for BBH mergers shifts to more equal mass ratios at the BH+O-star stage the higher the BH mass (Fig.~\ref{fig.appendix_4_7_10}), see also \citet{Picco2024}.

\textbf{BBH mass ratios.} We find BBH mergers formed with mass ratios from 0.3 to 1.0 (Fig.~\ref{fig.bbh_BRW}). These limits are determined by the requirement that SMT evolution in BH+O-star systems leads to short-period orbits. 
To form the secondary BH with mass more than 3 times the mass of the primary BH would require very wide BH+O-star systems, which do not follow the SMT evolution (convective donors). To form the secondary BH with mass less than 0.3 times the mass of the primary BH would mean that the SMT phase leads to orbits that are too wide. On the other hand, these conclusions are based on the assumption that the secondary BH forms in direct collapse with no mass loss. Taking uncertainty in the BH formation process into account, it is hard to put a robust lower limit on the mass ratios of BBH systems from the SMT channel. Nevertheless, we do not find any particular reason for a preference of BBH mass ratios between $\sim0.6-0.8$, as found in some population models \citep{Olejak2022,vanSon2022a}. However, this likely depends on the assumptions on the first mass transfer and possibly the metallicity-weighted cosmic star formation rate that we do not explore here \citep{vanSon2022a,vanSon2022b,Zevin2022,Banerjee2024,Dorozsmai2024}.

\textbf{BH spins.} Due to the separation limit, we find that BBH mergers from the SMT channel have orbital periods predominantly $\gtrsim 0.8$ days at the BH+He-star stage. These systems are generally too wide for an efficient tidal spin-up of the He-star, which has been the primary scenario explaining the origin of BH spins in the CE channel \citep{Kushnir2016,Qin2018,Bavera2020}. Even if tides were strong enough to fully synchronize the He-star with the orbit, for a period of $0.8$ day \citet{Belczynski2020} fit the secondary BH spin of only $\sim 0.2$. Therefore, our results suggest a limit on the maximum spin achievable in the SMT channel, which would impact the mass ratio reversal scenarios as an explanation for the reported anticorrelation between the effective spin ($\chi_{\rm eff}$) and mass ratio \citep{Broekgaarden2022b,Olejak2024,Banerjee2024}. On the other hand, the closest separation BH+He-star systems we find are preferentially of unequal mass ratio with $M_{\rm He;star} > M_{\rm BH;1}$ (Fig.~\ref{fig.fig3_dodgerblue}), originating from relatively wide pre-SMT orbits and CHeB donors (Fig.~\ref{fig.slice}). In some borderline cases their periods may extend down to $\sim 0.4$ day (Fig.~\ref{fig.before_and_after_L2}), which would correspond to $a_{\rm BH;2} \sim 0.5$ and BBH mergers with $\chi_{\rm eff} \sim 0.3-0.4$ and $q_{\rm BBH}$ down to $0.3$.
More generally, however, without strong tides at the BH+He-star stage, the rotation of the second BH progenitor may predominantly be determined by the SMT phase itself \citep{Qin2019}, potentially leading to notable spins even in $>1$ day orbits. A more in-depth analysis of BH spins in the SMT channel is beyond the scope of this paper.

\textbf{Metallicity-dependence.} We found long delay times of BBH mergers at the favorable, low metallicity of $Z = 0.1\Zsun$. At higher metallicity, the winds of (He-)stars will be stronger, likely leading to more significant orbital expansion at the BH+He-star stage and even longer delay times. 
Our limited exploration of the metallicity effects suggests that this may quench the SMT channel completely at $Z \gtrsim 0.4\Zsun$ (Fig.~\ref{fig.appendix_metallicity}). 
This, however, is sensitive to the uncertain strength of winds, particularly from He-rich stars \citep{Vink2017,Sander2020a,Sander2020b,Vink2022araa,Dorozsmai2024}, as well as the orbital evolution due to wind-mass loss \citep[e.g.][]{Schroder2021}.

\textbf{Rate calculation.} 
Our results suggest that the parameter space for the formation of BBH mergers from BH+O-star systems is only weakly affected by mass transfer uncertainties (Fig.~\ref{fig.before_and_after_L2}). 
This means that by knowing how many BH+O-star systems form in the Universe, it may be relatively 
straight forward \citep[compared to many other formation scenarios][]{MandelBroekgaarden2022} to accurately obtain the rate of BBH mergers from SMT evolution.
While so far only a few BH+O-star systems are known, this number is set to increase substantially in the following years thanks to large-scale astrometric, spectroscopic, and microlensing surveys of tens of thousands of massive stars \citep{ShenarBodensteinerBloem2024,ElBadry2024NewAR,Abrams2025}, predicted to yield tens to hundreds of BH binary detections \citep{Wiktorowicz2020,Langer2020,Janssens2022}.

\subsection{Importance of mass transfer stability for the SMT channel and how to model it in rapid binary codes}

\label{sec.disc_stability_treatment}

Criteria for mass transfer stability are, arguably, the single most important assumption to make when modeling the formation of compact binary systems, such as merging BBHs, through SMT evolution. This is because close orbital separations are reached via SMT only in systems that evolve near the threshold between stable and unstable MT evolution (e.g. Fig.~\ref{fig.grid_analyt_vs_MESA}). As a result, it is the stability threshold that determines the parameter space for the formation of BBH merges via SMT and consequently the BBH merger properties (masses, mass rations, delay times). For example, because the critical mass ratio for stability increases with the radius of the radiative donor (Fig.~\ref{fig.triumvirate}), the mass ratios of BBH mergers correlate with the orbital period at the BH+Star stage (Fig.~\ref{fig.bbh_BRW}). Stability of mass transfer across different donors determines the mass range of BH progenitors in which the SMT channel operates (Fig.~\ref{fig.obywatel_gc}). The fundamental separation limit responsible for long delay times of BBH mergers is caused by mass transfer stability requirements (Sec.~\ref{subsec.res_consequences_BBHmergers}). 
The few cases of population synthesis studies that have explored different assumptions on stability have indeed found a substantial effect on predictions from the SMT channel \citep{Olejak2021,ShaoLi2021,vanSon2022b,Dorozsmai2024}.

The question arises how to best model mass transfer stability in rapid binary codes to capture the results and trends found in detailed models. A common approach \citep[e.g.][]{Hurley2002,Claeys2014,Breivik2020,Riley2022_compas} is to determine stability based on several fixed critical mass ratio values (or equivalent mass-radius exponents) for stars at different evolutionary stages, with a possible variation as a function of the core mass fraction in convective giants \citep{Webbink1985,Hjellming1987}. We find that this is strongly inconsistent with stability in detailed stellar models, which indicate a clear trend of $q_{\rm crit}$ increasing with radius for radiative donors (e.g. $q_{\rm crit}$ ranging from $2$ to $7$ for a $30\Msun$ donor depending on its radius, see Fig.~\ref{fig.triumvirate}). While here we show it with full 1D stellar models, this point was made and discussed already by \citet{Ge2015} and later \citet{Ge2020}, using their adiabatic mass-loss model \citep{Ge2010}. 

Here, we propose a revised approach for the treatment of mass transfer stability in population synthesis, motivated by the results in Sec.~\ref{subsec.res_it_is_the_star_that_matters}. 
Rather than deciding based on the mass ratio at RLOF, we suggest to determine stability based on the minimum orbital separation that may be reached via SMT evolution. Namely, if at the end of an SMT phase the orbit was to become smaller than the separation limit of the donor, the mass transfer becomes unstable. 
This approach is attractive for several reasons: (a) the minimum separations achievable from SMT evolution for different donors are robust against uncertainties of mass transfer (accretion efficiency, AM loss), whereas critical mass ratios are not (Fig.~\ref{fig.triumvirate}); (b) for radiative donors, it is the orbital separation near the end of the SMT phase that is directly connected to the physical origin of instability (Sec.~\ref{subsec.disc_why_the_limit_exists}); (c) for a given donor type, the minimum separation from SMT varies more slowly with the donor radius than the critical mass ratio; (d) final orbital separation after mass transfer is often the most relevant quantity for an evolutionary channel (e.g. formation of BBH mergers).
Furthermore, in Sec.~\ref{sec.disc_why_donors_differ} we discuss how the minimum separation from SMT evolution can be related to the internal structure of the donor star.  This opens an exciting new avenue of using the population of GW sources to constrain models of stellar interiors and solve puzzles such as the blue supergiant problem (Sec.~\ref{sec.disc_GW_stellar_interiors}). However, this only becomes possible if the separation limit is self-consistently implemented in the treatment of SMT evolution. Even having the most precise mapping of critical mass ratios for donors of different mass and radius will result in altered post-SMT separations when used across different binary evolution codes (e.g. due to the amount of mass transfer code-different). 

The minimum orbital separation from SMT evolution obtained in this work are available on Zenodo.\footnote{Upon acceptance of the paper. Currently available on request.} As a rule of thumb, minimum separation from MS donors are between $\sim15$ and $30\Rsun$, for HG donors are between $\sim30$ and $50\Rsun$, and for CHeB donors are between $\sim8$ and $15\Rsun$ (all for radiative-envelope donors). For convective-envelope donors, the stability is better described with critical mass ratios or mass-radius exponents.

\section{Summary}
\label{sec.summary}

At the advent of a rapid increase in the number of detections of compact binary mergers, theoretical predictions of the population of GW sources are highly needed and yet, still, highly uncertain. 
A promising formation scenario, particularly for binary BH mergers, is stable mass transfer evolution in massive binary systems. Its main open questions have to do with how the binary orbit evolves during mass transfer under the influence of complex gas dynamics of high-mass X-ray binaries. Here, we studied the SMT channel trough detailed binary evolution computations of interacting systems with a BH accretor and a massive star companion (BH+O-star), assisted with analytical results for interpretation. We find that:

\textbf{(a)} There is a fundamental limit to how close binary systems can get via SMT evolution, that is robust against uncertainties in binary physics of orbital evolution. 
Even under the assumption of very efficient orbital shrinkage due to angular momentum loss through L2 outflows, the final size of the orbit is always larger than $\gtrsim$10$\Rsun$. Systems evolving towards tighter orbits become dynamically unstable and result in stellar mergers instead (Fig.~\ref{fig.fig3_dodgerblue}).
While it remains uncertain which BH+Star systems evolve to become GW sources due to mass transfer unknowns, the outcomes of the SMT channel all converge to the same narrow window of final orbits of merging BBH systems across different model variations (Fig.~\ref{fig.before_and_after_L2}), which facilitates confident predictions for the properties of BBH mergers. 

\textbf{(b)} The orbital separation limit has direct implications for the properties of BBH mergers in the SMT channel: long delay times ($\gtrsim$1-2 Gyr) and a lack of a high BH spins from the tidal spin-up of their helium star progenitors (Fig.~\ref{fig.bbh_delaytimes} and Sec.~\ref{sec.disc_gw_implications}). At high metallicity, where the post-mass transfer orbits are expected to further expand due to strong Wolf-Rayet winds, the SMT channel may become severely quenched. (Fig.~\ref{fig.appendix_metallicity}). We find BBH mergers with masses from $10\Msun$ to $90\Msun$ and mass ratios from $0.3$ to $1$ (Fig.~\ref{fig.bbh_BRW}). Case A mass transfer dominates the formation of massive BBH mergers ($\gtrsim40\Msun$).

\textbf{(c)} The reason for the separation limit lies in the properties of the stellar structure (as opposed to in binary physics), which dictates when a system will become unstable. If the orbit gets too narrow during mass transfer, a delayed dynamical instability is triggered by a rapid expansion of the remaining donor envelope due to its near-flat entropy profile (Fig.~\ref{fig.disc_point_prenvelopeofiles} and Sec.~\ref{subsec.disc_why_the_limit_exists}). The orbital size when this may occur depends on the entropy structure of the donor, which varies across evolutionary types. The closest separation from SMT evolution can be achieved from core-He burning donors ($\sim$8$-$15$\Rsun$) and late Main sequence donors ($\sim$15$-$30$\Rsun$), while Hertzsprung Gap donors lead to wider orbits ($\gtrsim$30$-$50$\Rsun$) and wide non-merging BBHs (Fig.~\ref{fig.disc_different_donors}).

\textbf{(d)} Close-orbit binaries from the SMT channel (such as BBH mergers or stripped-star systems) can be a unique probe of chemical mixing in stars. This is because any chemical abundance gradient in a radiative envelope of the donor introduces an entropy gradient, which stabilizes the mass transfer and allows for a tighter final orbit. Extended regions of constant composition (e.g. inner convective zones) have the opposite effect. The same internal mixing in stars that is important for the blue-supergiant problem may have a substantial effect on the formation of BBH mergers (Fig.~\ref{fig.obywatel_gc} and Sec.~\ref{sec.disc_GW_stellar_interiors}).

Our findings help to identify the fundamental predictions of the SMT channel that may be used to discriminate which formation scenarios contribute to the observed population of GW sources. This will become particularly important in the near future, when the growing dataset of GW sources allows for the measurement of trends with redshift and spin. To incorporate our results into population models of GW sources, we propose a new treatment of mass transfer stability for rapid binary codes (Sec.~\ref{sec.disc_stability_treatment}).

\begin{acknowledgements}
This research was supported by the Munich Institute for Astro-, Particle and BioPhysics (MIAPbP) which is funded by the Deutsche Forschungsgemeinschaft (DFG, German Research Foundation) under Germany´s Excellence Strategy – EXC-2094 – 390783311. This research was supported by the International Space Science
Institute (ISSI) in Bern, through ISSI International Team project {\#}512
(''Multiwavelength View on Massive Stars in the Era of Multimessanger Astronomy'').
This work was performed in part at Aspen Center for Physics, which is supported by National Science Foundation grant PHY-2210452. AO acknowledge funding from the Netherlands 732 Organization for Scientific Research (NWO), as part of 733 the Vidi research program BinWaves (project number 734 639.042.728, PI: de Mink).
\end{acknowledgements}

\bibliographystyle{aa}
\bibliography{ULX_bib.bib}

\begin{appendix}

\section{Semi-analytical model for stable-mass transfer evolution}

\label{app.semianalytical_model}

Here we describe our method to estimate the outcome of an SMT phase in a BH+Star binary (used, for instance, to obtain the orbital evolution in Fig.~\ref{fig.analyt_MT_sketch}). The goal is not have a fully accurate model but rather a simple tool to compare our expectations with the results from detailed binary simulations. 

We consider a binary with a BH $M_{\rm BH}$ and a star $M_{\rm star;ini}$ that initiates mass transfer when the orbital separation is $a_{\rm ini}$ (and the eccentricity is zero). The change of separation as a result of an SMT phase can be written as:
\begin{equation}
    a_{\rm fin} = a_{\rm ini} + \int_{M_{\rm star;ini}}^{M_{\rm star;final}} \frac{{\rm d}a}{{\rm d}M_{\rm star}} \left( M_{\rm BH}, \beta, f_{\rm wind}, f_{\rm bipolar}, f_{\rm L2} \right)    {\rm d}M_{\rm star} \, \, ,
\label{eq.app_SMT_integral}
\end{equation}
where the expression for ${\rm d}a / {\rm d}M_{\rm star}$ can be found by combining Eqn.~\ref{eq.da_a} and Eqn.~\ref{eq.gamma}. To compute the integral in Eqn.~\ref{eq.SMT_integral}, we need to know $M_{\rm BH}$, $\beta$, $f_{\rm wind}$, $f_{\rm bipolar}$, $f_{\rm L2}$, and $M_{
\rm star;final}$. The BH mass $M_{\rm BH}$ is given by the initial conditions, alongside $a_{\rm ini}$ and $M_{\rm star;ini}$. We assume that the accretion rate onto the BH is Eddington-limited, which in the case of BH+Star systems with massive star donors leads to nearly fully non-conservative mass transfer with $\beta \approx 0$ and $M_{\rm BH} \approx {\rm const}$ \citep[e.g.][]{Klencki2019}. Furthermore, the fraction of the mass lost via wind is:
\begin{equation}
   f_{\rm wind} = \frac{\Delta M_{\rm star;wind}}{M_{\rm star;ini}-M_{\rm star;final}} = \frac{ \dot{M}_{\rm star;wind}\Delta t_{\rm SMT}}{\Delta M_{\rm star}} \, \, \, ,
\end{equation}
where $\dot{M}_{\rm star;wind}$ is the average mass-loss rate via stellar wind and $\Delta t_{\rm SMT}$ is the duration of the SMT phase. Fractions $f_{\rm bipolar}$ and $f_{\rm L2}$ are free parameters that need to satisfy $f_{\rm bipolar} + f_{\rm L2} = 1 - f_{\rm wind}$.

To gauge how much mass is lost from the donor ($\Delta M_{\rm star}$) and on what timescale ($\Delta t_{\rm SMT}$), we take a look at the evolutionary stage of the donor star. The first possibility is that the donor is a post-Main Sequence (post-MS) giant star, with a helium core $M_{\rm core}$ and a hydrogen envelope $M_{\rm env}$, i.e. case B or case C mass transfer \citep{Kippenhahn1967}. In that case, we assume $\Delta M_{\rm star} = 0.9 M_{\rm env}$, i.e. the mass transfer strips nearly the entire envelope (93\%). The exact choice of 93\% is somewhat arbitrary but it is motivated by our detailed binary models of BBH merger progenitors in which the remaining unstripped envelope fraction is typically between 5\% and 10\% \citep[in SMT leading to wide orbit BBH systems, the amount of envelope left unstripped can be much larger,][]{Klencki2022}. We further assume thermal-timescale mass transfer rate $\dot{M}_{\rm star} = 10^{-3}\Mpy$, such that the duration $\Delta t_{\rm SMT} = \Delta M_{\rm star} / \dot{M}_{\rm star}$ is typically of the order of $10^4\yr$.

In the case of a donor that is a MS star (case A interaction), there is no distinct core-envelope structure yet to estimate the amount of mass that is transferred. In general, $\Delta M_{\rm star}$ will depend on the parameters of each individual system such as the orbital period, mass ratio, and the donor mass \citep[see the analytical fits and the methods by][]{Schurmann2024, Shikauchi2024}. Case A mass transfer is also typically not a single short-lived event but rather a series of two or more separate interaction phases, some happening on the nuclear timescale, taking place all the way until the end of the MS \citep{Pols1994,Wellstein2001,MarchantBodensteiner2023}. Right afterwards, the post-MS expansion of the remaining donor leads to another interaction referred to as case AB mass transfer, that further strips the star of mass. Here, for simplicity, we model this entire sequence of interactions (case A + AB) as a single SMT phase with the duration $\Delta t_{\rm SMT}$ equal to remaining MS lifetime from the onset of RLOF. We further assume that the final mass of the donor is $M_{\rm star;final} = M_{\rm star;TAMS} - 0.9 M_{\rm env;TAMS}$, 
where $M_{\rm star;TAMS}$ and $M_{\rm env;TAMS}$ are respectively the mass of the donor and the mass of its envelope at terminal-age MS (TAMS) if the donor was to continue its evolution without ever engaging into mass transfer (as a single star). This approach is quite accurate for case A mass transfer initiated during an advanced MS stage ($X_{\rm center} \lesssim 0.2-0.3$) but it underestimates the amount of mass that is transferred in very-close binaries that interact earlier. 

To obtain $M_{\rm star;final}$ and $f_{\rm wind}$, we thus need to know several parameters of the donor star: its evolutionary stage, core mass, remaining MS lifetime (for MS stars), and the wind mass-loss rate. Those could be derived from analytical fits to stellar tracks by \citet{Hurley2002}, as done in various population synthesis codes. Here, we use our custom and publicly-available massive star models from \citet{Klencki2020}, with more up-to-date treatment of mixing and mass loss. 

Although simplistic, the mass-transfer model presented above is similar to how SMT evolution (in particular case B/C interaction) is treated in rapid binary-evolution codes widely used for GW-source population synthesis.\footnote{One key additional component is the criteria for mass transfer stability, which greatly vary across different codes and their iterations.} 

\section{Uncertainty: BH natal kicks}
\label{sec.disc_kicks}

Throughout the paper, we have assumed that the secondary BH forms via direct collapse with no natal kick. While tentative evidence suggests that complete collapse may occur in stellar-mass black holes \citep{2003Sci...300.1119M,2022NatAs...6.1085S,2024PhRvL.132s1403V}, the conditions required are not fully understood. 
Moreover, even in a complete collapse scenario, uncertainties remain regarding the amount of mass radiated in neutrinos and the natal kick they impart. From simulations, the mass decrement from neutrino emission is less than $0.6 M_{\odot}$ \citep[e.g.,][]{2021ApJ...909..169K} and the neutrino natal kick magnitude is likely 
% \avg{[given than Burrows group says it can be as large as 70 km/s or something like that, I can find the number]}
around a few or a few tens of km/s \citep[see][and references therein]{2024Ap&SS.369...80J}.
Recently, \cite{2024PhRvL.132s1403V} used VFTS 243 to constrain the neutrino natal kick magnitude to $\lesssim 10$ km/s, corresponding to neutrino emission asymmetries of $\lesssim 4\%$. On the other hand, space distribution and proper motions of Galactic BH low-mass X-ray binaries point towards BH kicks with magnitudes up to $\sim100$ km/s \citep{Repetto2012,Repetto2015,Mandel2016}. 

Altogether, these findings suggest that while the natal kicks of BHs are likely not as strong as in the case of NSs, our default assumption of a zero kick and no mass loss may be too simplistic. The main question is how much may it affect the delay times of BBH mergers obtained in our study. In the case of mass loss with no additional kick, a spherically symmetric mass ejection would make the orbit wider everywhere except at the periastron \citep{1961BAN....15..265B}, thus weakening 
% \avg{[Hmn, uncertain if weakening is the right word]}
the GW emission and always leading to longer delay times. If the newly-formed black hole receives an additional velocity kick, the impact on its orbit will depend on the direction of the kick. 
% If the newly-formed BH received an additional velocity kick, its effect on the orbit will depend on the kick direction, as discussed in \citet{Brandt1995} for X-ray binaries and \citep{Marchant2016} in the context of binary black hole mergers, and in great detail in Vigna-G\'omez (in prep.)
This has been discussed in the literature in the context of X-ray binaries \citep{Brandt1995}, binary black hole mergers \citep{Marchant2016}, and recently in the context of BBHs formed via the complete collapse scenario (Vigna-G\'omez in preparation). 
Broadly, the resulting BBH delay time becomes notably shorter only if the kick is directed in the opposed direction to the orbital movement and its magnitude is comparable to the orbital velocity, which in the case of BH + massive He stars systems are of the order 200-400 km/s.  For example, for isotropic BH kicks with the magnitude equal to half the orbital velocity (i.e. 100-200 km/s), in 20\% of cases the BBH delay time would become shorter by a factor of at least 10 (Vigna-G\'omez in preparation.). BH kicks of a few tens of km/s or less would have a negligible effect on the resulting BBH delay times.

\section{Typo in the angular momentum loss through L2 in MESA models}
\label{app.L2_mishap}

\begin{figure}[h]
    \includegraphics[width=\columnwidth]{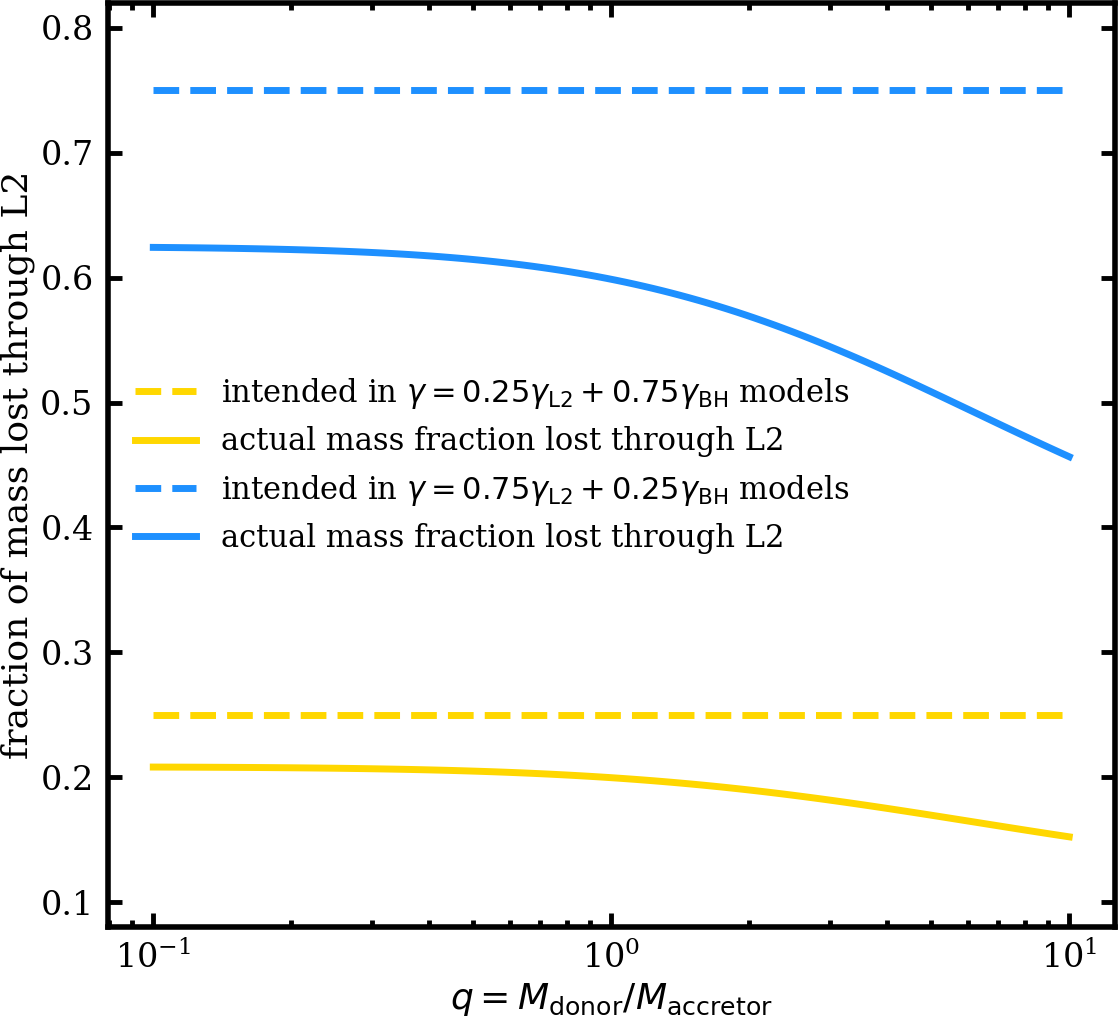}
    \caption{Comparison of the intended  versus the actual effective fraction of the mass lost through L2 in our MESA models (see text for explanation of the discrepancy). The analytical predictions of the orbital separation during SMT in Fig.~\ref{fig.grid_analyt_vs_MESA} follow the actual mass loss fractions, i.e. the solid lines, so that the they could be directly compared with the MESA models.}
    \label{fig.app_L2_mishap}
\end{figure}

Throughout the paper, we consider three modes of ejection of the non-accreted matter in an interacting BH + Star binary: with the wind of the donor, as a bipolar outflow or fast wind launched close to the BH accretor, or as mass ejected from the proximity of the L2 point. Each case corresponds to different specific amount of angular momentum that is lost, parametrized by $\gamma$ (see Sec.~\ref{subsec.method_orbit_equations} for details). In particular, for the mass loss through the L2 point, we recall:
\begin{equation}
\gamma_{\rm L2} = \Big( \frac{a_{\rm L2}}{a}  \Big)^2 \frac{(M_{\rm BH}+M_{\rm star})^2}{M_{\rm BH}M_{\rm star}} \approx 1.44 \frac{(M_{\rm BH}+M_{\rm star})^2}{M_{\rm BH}M_{\rm star}}
\end{equation}
where $a_{\rm L2}$ is the distance from the L2 to the center of mass that for most mass ratios can be approximates as $a_{\rm L2} \approx 1.2 a$ \citep{Pribulla1998}. The above equation assumes that at the location of L2 the ejected mass co-rotates with the binary. 

In the binary module of MESA \citep{Paxton2015}, the treatment of the non-accreted matter is set up with several \texttt{mass\_transfer} controls. In particular, \texttt{mass\_transfer\_delta} defines the fraction of mass that is lost via a circumbinary coplanar toroid. The radius of the toroid $a_{\rm cct}$ is set by the \texttt{mass\_transfer\_gamma} parameter, such that $(a_{\rm cct} / a)^{0.5}=$ \texttt{mass\_transfer\_gamma}. Crucially, the toroid is assumed to rotate with its own period of a Keplerian orbit, rather than to corotate with the binary. As a result, the specific angular momentum of the mass lost from the toroid can be characterized by $\gamma_{\rm CCT}$:
\begin{equation}
\gamma_{\rm CCT} = \sqrt{\frac{a_{\rm CCT}}{a}} \frac{(M_{\rm BH}+M_{\rm star})^2}{M_{\rm BH}M_{\rm star}} \; \; \; \; \; .
\end{equation}

That means to obtain $\gamma_{\rm CCT} = \gamma_{\rm L2}$ and model mass loss through the L2 point in MESA, one needs to set \texttt{mass\_transfer\_gamma}$=1.44$ and 
\texttt{mass\_transfer\_delta} to the intended fraction of mass lost through L2.
In our case, we have set \texttt{mass\_transfer\_delta} = 0, 0.25, and 0.75 (depending on the model variation) and, unfortunately, \texttt{mass\_transfer\_gamma}$=1.2$ instead of $1.44$. This means that the amount of AM lost with L2 outflows was lower than intended. Alternatively, this means that the effective fraction of mass lost through L2 in our models was lower than the intended value set with \texttt{mass\_transfer\_delta}. The correction factor depends on the mass ratio, as illustrated in Fig.~\ref{fig.app_L2_mishap}. Effectively, in MESA models with \texttt{mass\_transfer\_delta} = 0.25 and \texttt{mass\_transfer\_gamma} = 1.2, the actual fraction of mass lost through L2 was about $\sim0.2$. We refer to them as enhanced orbital shrinkage in the main text, with $\gamma \approx 0.2 \gamma_{\rm L2} + 0.8 \gamma_{\rm BH}$. Similarly, in models with \texttt{mass\_transfer\_delta} = 0.75 and \texttt{mass\_transfer\_gamma} = 1.2, the actual fraction of mass lost through L2 was about $\sim0.55$. In the main text these are referred to as extreme orbital shrinkage with $\gamma \approx 0.55 \gamma_{\rm L2} + 0.4 \gamma_{\rm BH}$ (Fig.~\ref{fig.before_and_after_L2}). Fig.~\ref{fig.app_L2_mishap} reveals that as the mass ratio $M_{\rm donor}/M_{\rm accretor}$ decreases during a mass transfer phase, the effective fraction of L2 mass loss in our models slowly increases. We do not expect this to have any effect on the conclusions presented in the paper and the predicted existence of a separation limit from SMT evolution. If anything, the enhanced L2 mass loss at late stages of an interaction (after the constant entropy layers have already been stripped, e.g. MS or HG donors) might allow for a somewhat smaller final separation compared to models with a constant AM loss. 

\section{Additional Figures}
\label{sec.app_addFigs}

\begin{figure*}
    \includegraphics[width=0.85\textwidth]{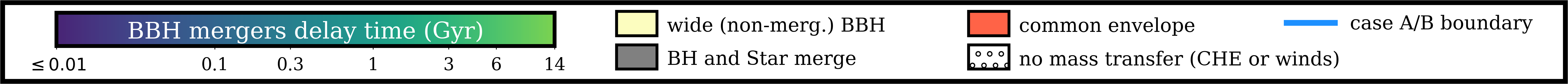}
    \includegraphics[width=0.85\textwidth]{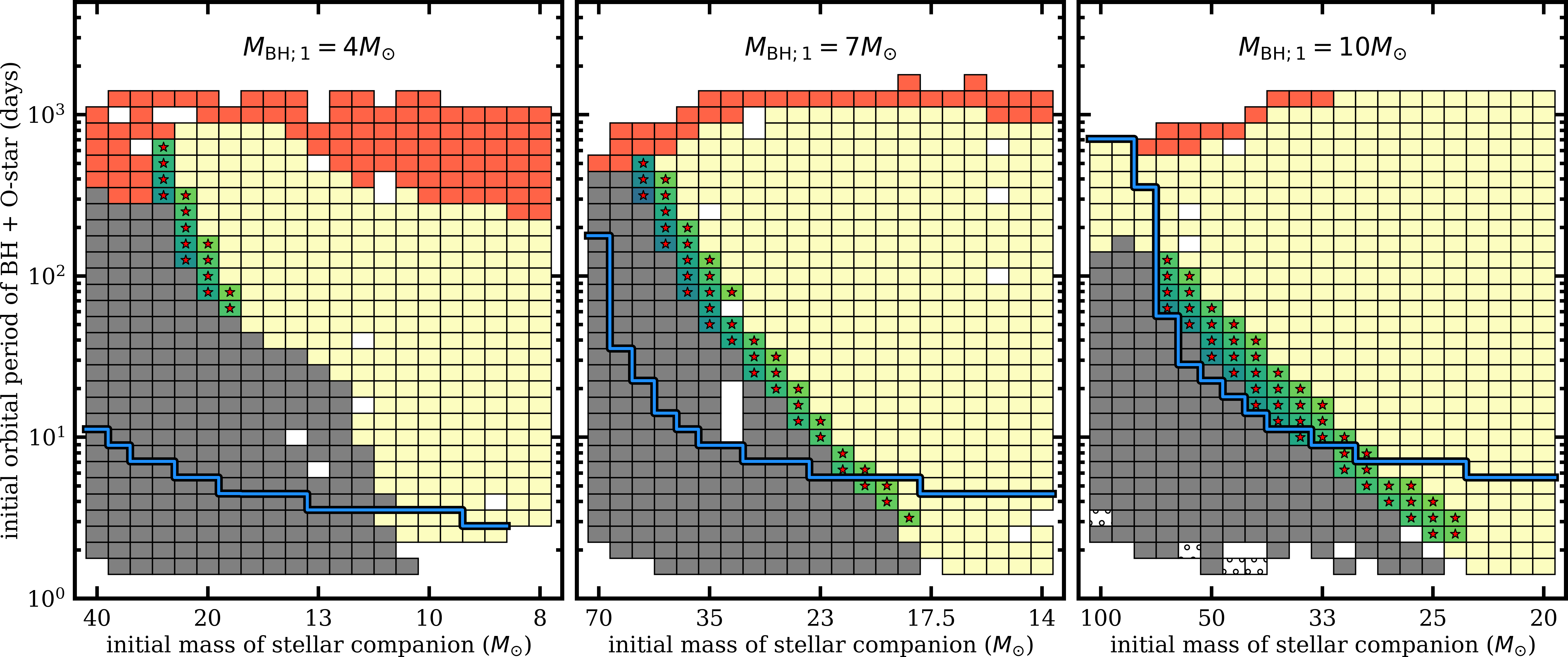}
    \includegraphics[width=0.85\textwidth]{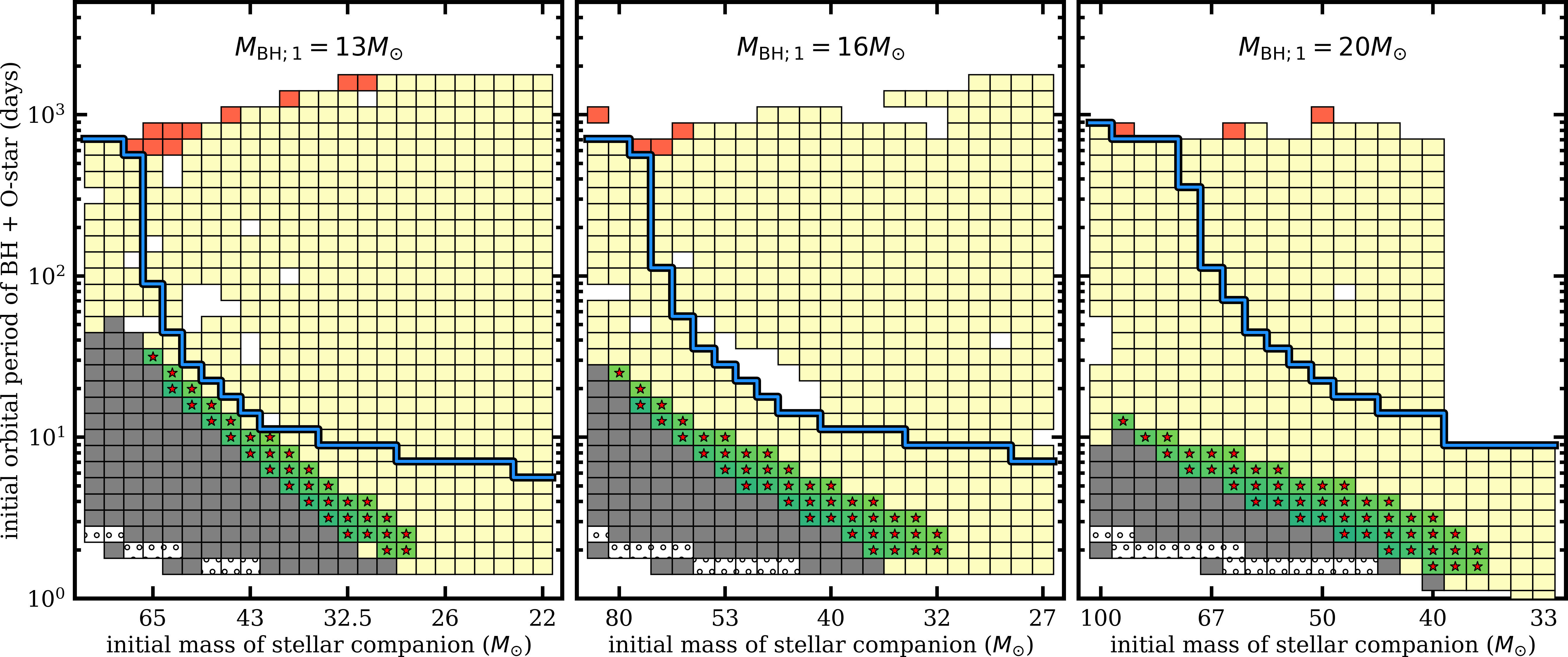}
    \includegraphics[width=0.85\textwidth]{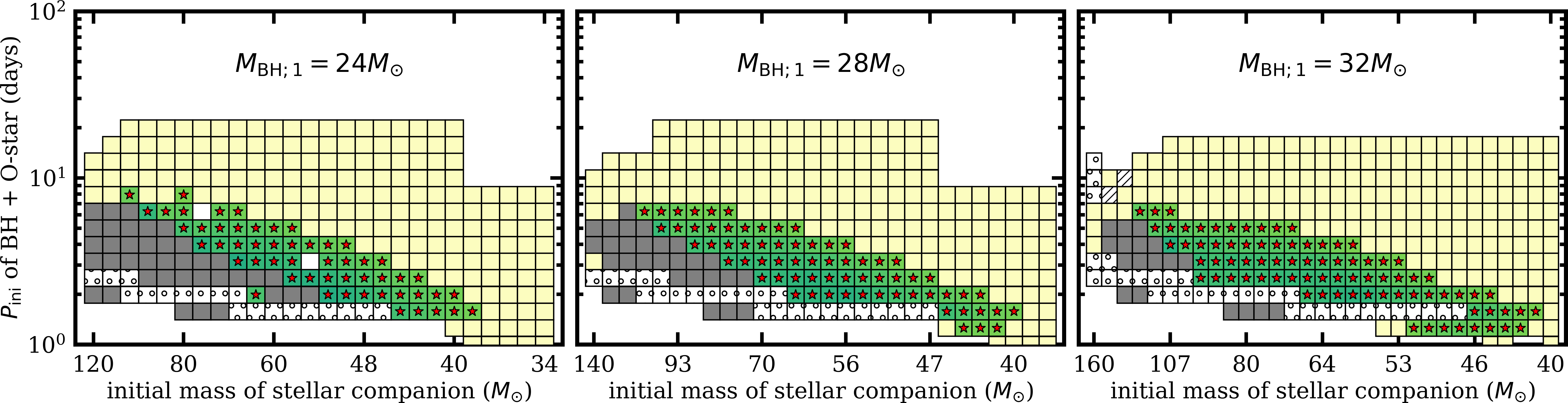}
    \includegraphics[width=0.578\textwidth]{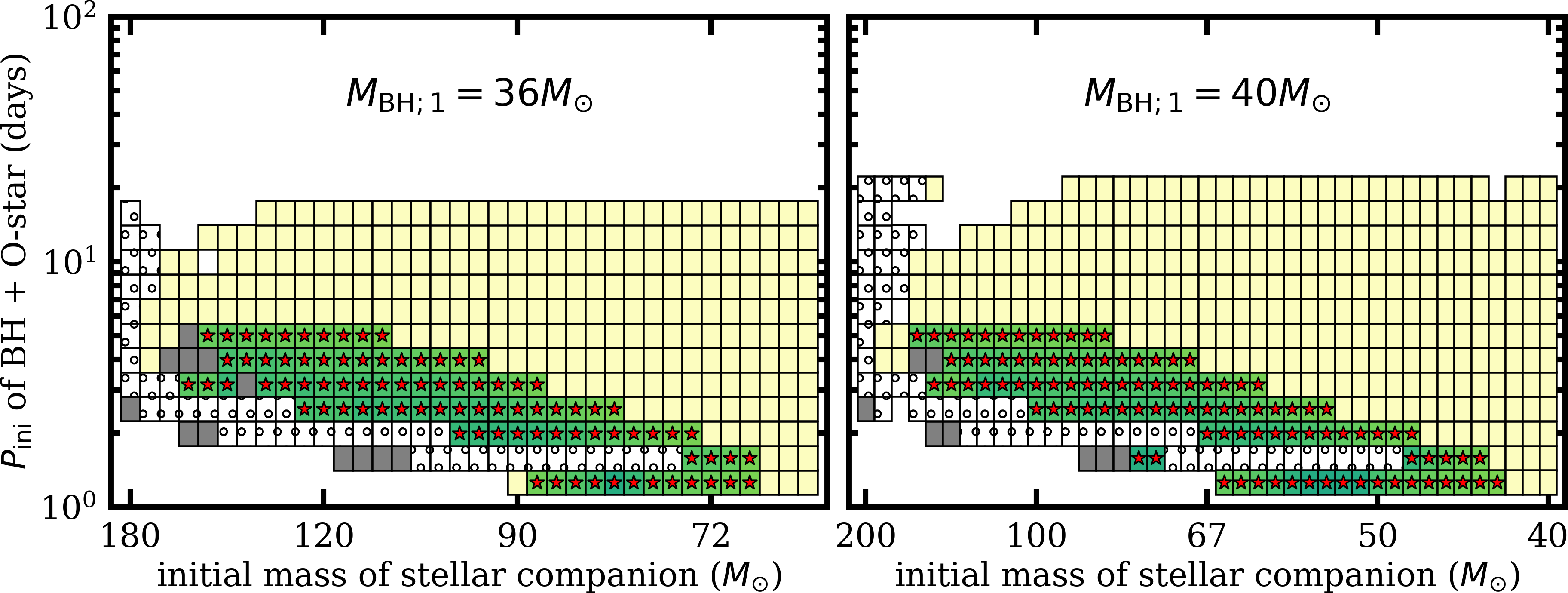}
    \caption{The parameter window for BBH mergers from SMT evolution across different BH masses. Here, we plot the outcome of binary MESA models marking BBH mergers with star symbols, assuming fiducial AM loss ($\gamma = \gamma_{\rm BH}$) and different primary BH masses $M_{\rm BH;1}$. Color indicates the final fate: wide BBH system (light yellow), merging BBH system (colorscale indicating the delay time), unstable MT with a convective donor (red), unstable MT with a radiative donor, likely leading to a BH + star merger (grey), close non-interacting systems due to chemically-homogeneous mixing or very strong winds (white dotted). In the top two rows, the blue line marks the boundary between MS and post-MS donors. In the bottom two rows, only a limited range of periods is computed and all the donors are on the MS. The grid is constructed based on the initial mass ratio $M_{\rm BH;1} / M_{\rm star}$ in steps of $0.02$. 
}
    \label{fig.appendix_4_7_10}
\end{figure*}

\begin{figure*}
    \includegraphics[width=0.85\textwidth]{Figures/mtstab_bbhM1_firstplot_v4_legend.png}
    \includegraphics[width=0.85\textwidth]{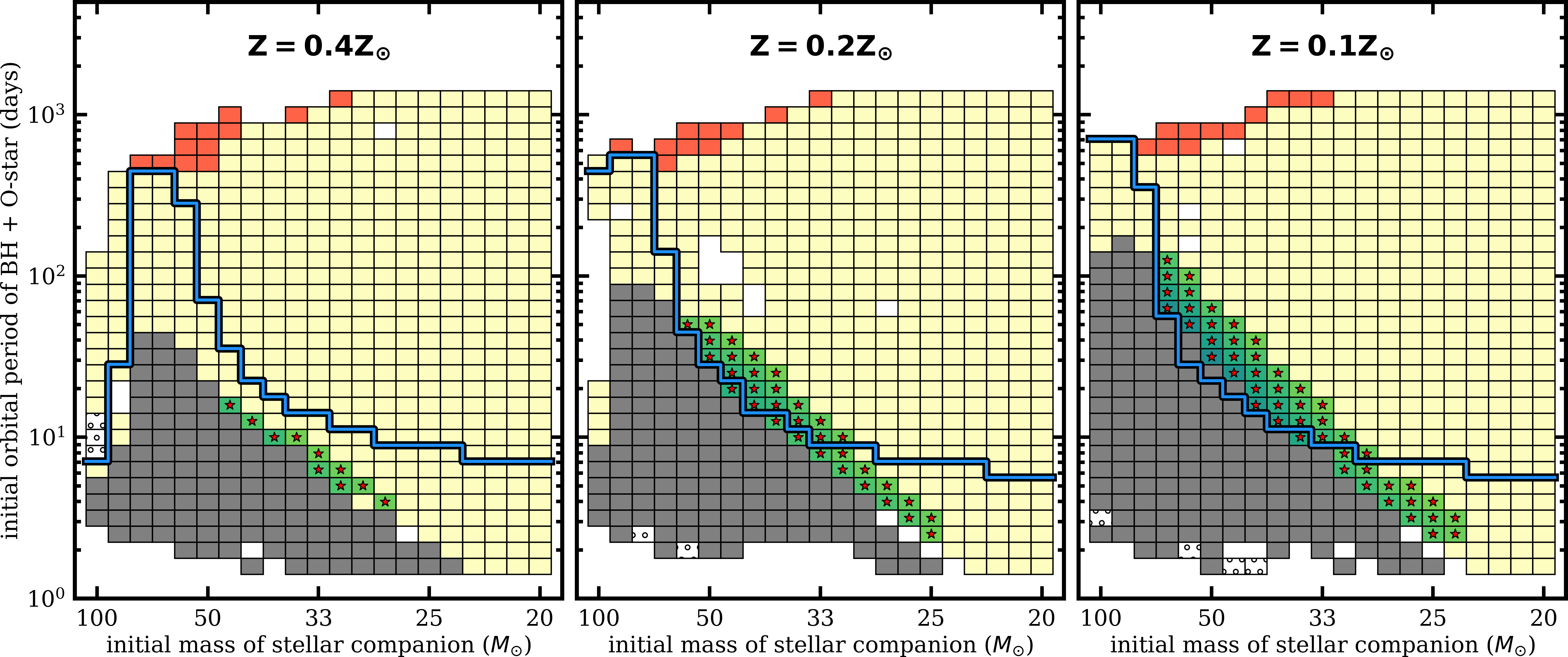}
    \includegraphics[width=0.578\textwidth]{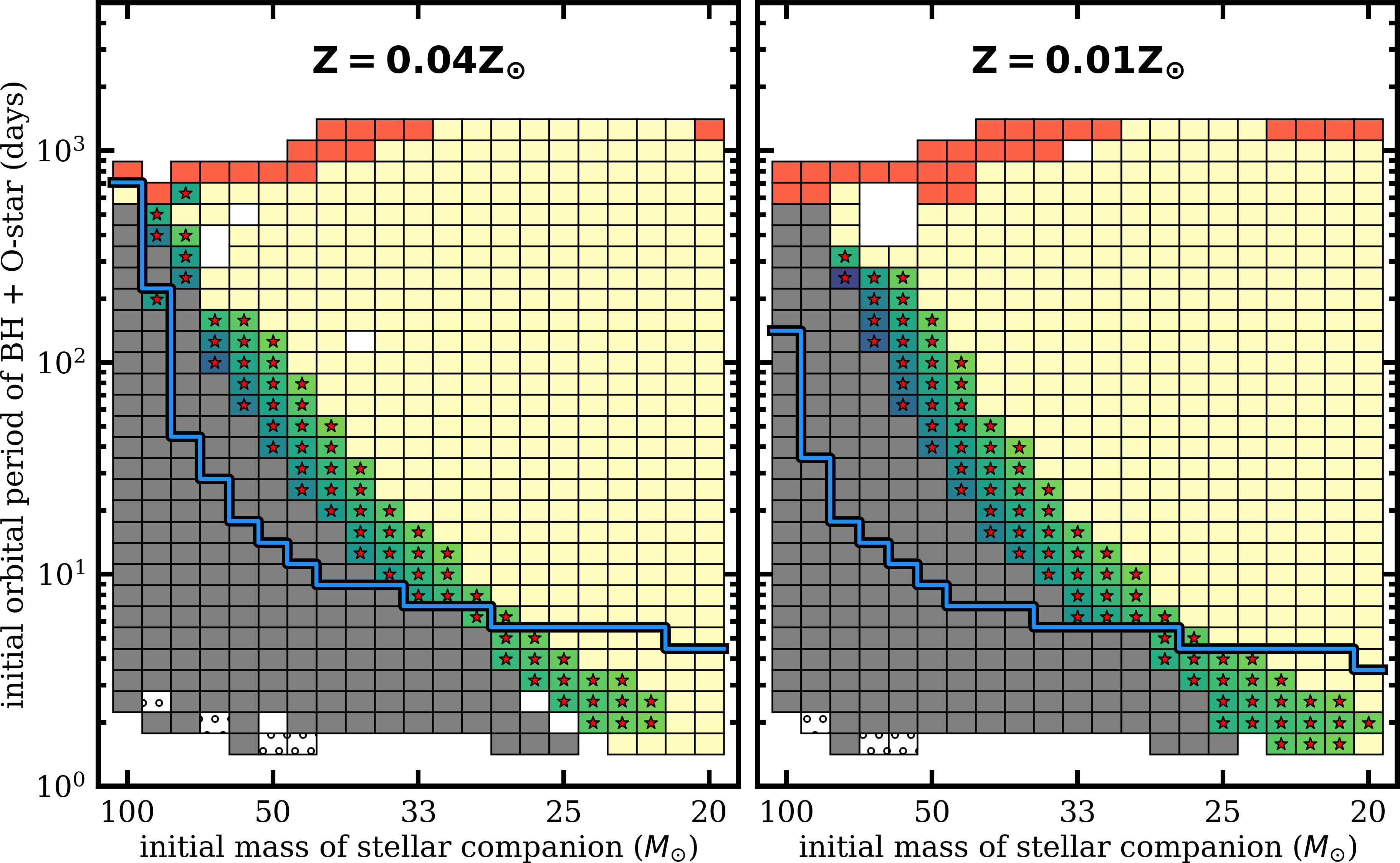}
    \caption{The SMT channel may become ineffective at high metallicity because of the combined effect of the separation limit in SMT evolution and orbital widening via Wolf-Rayet winds. Wwe plot the outcome of binary MESA models at different metallicities, marking BBH mergers with stars. The first BH mass is $M_{\rm BH;1} = 10 \Msun$ and we assume the fiducial AM loss for SMT ($\gamma = \gamma_{\rm BH}$). Color indicates the final fate, same as Fig.~\ref{fig.appendix_4_7_10}. The blue line marks the boundary between MS and post-MS donors.
    }
    \label{fig.appendix_metallicity}
\end{figure*}

\begin{figure*}
    \includegraphics[width=0.9\columnwidth]{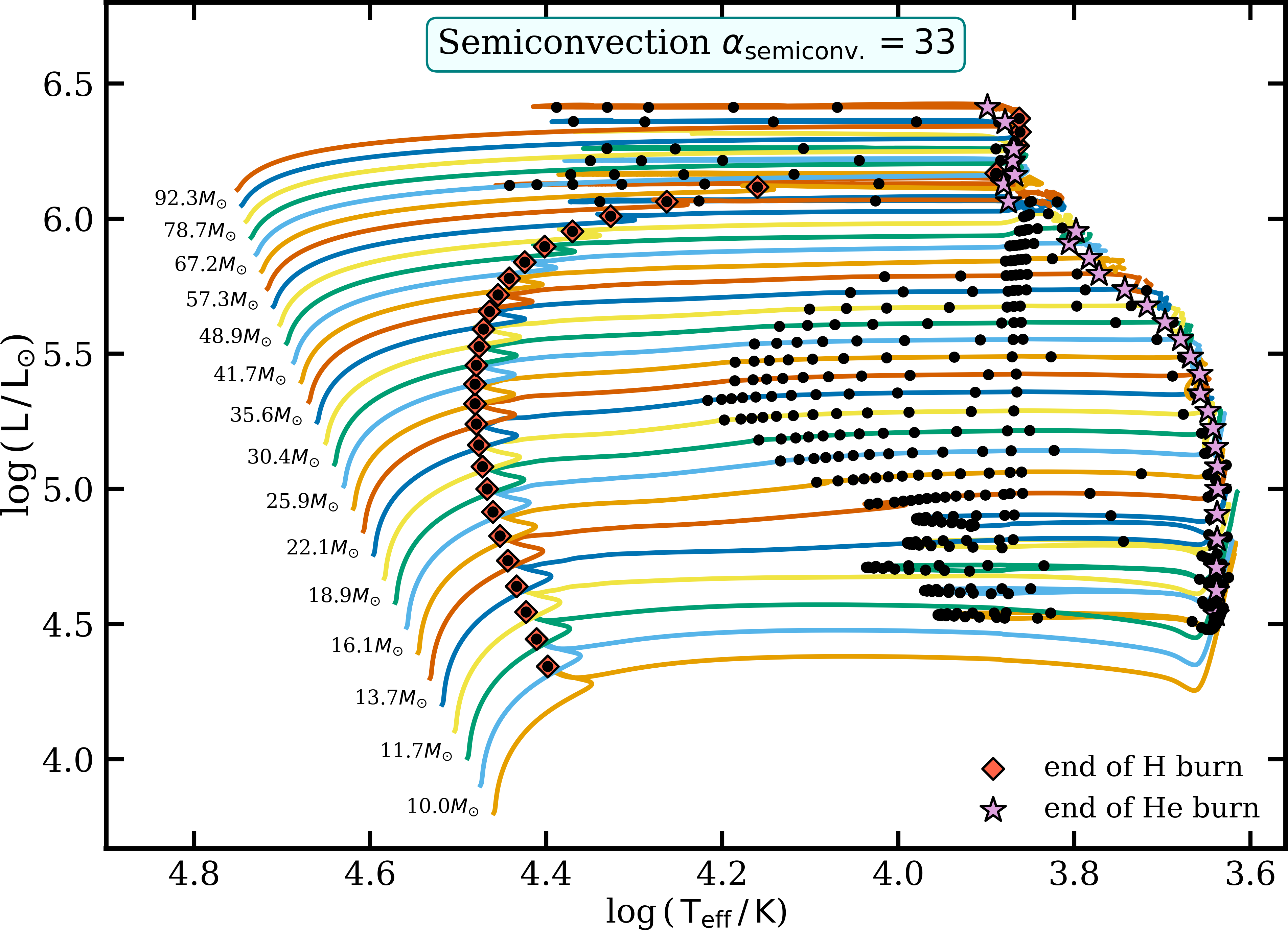}
    \includegraphics[width=0.9\columnwidth]{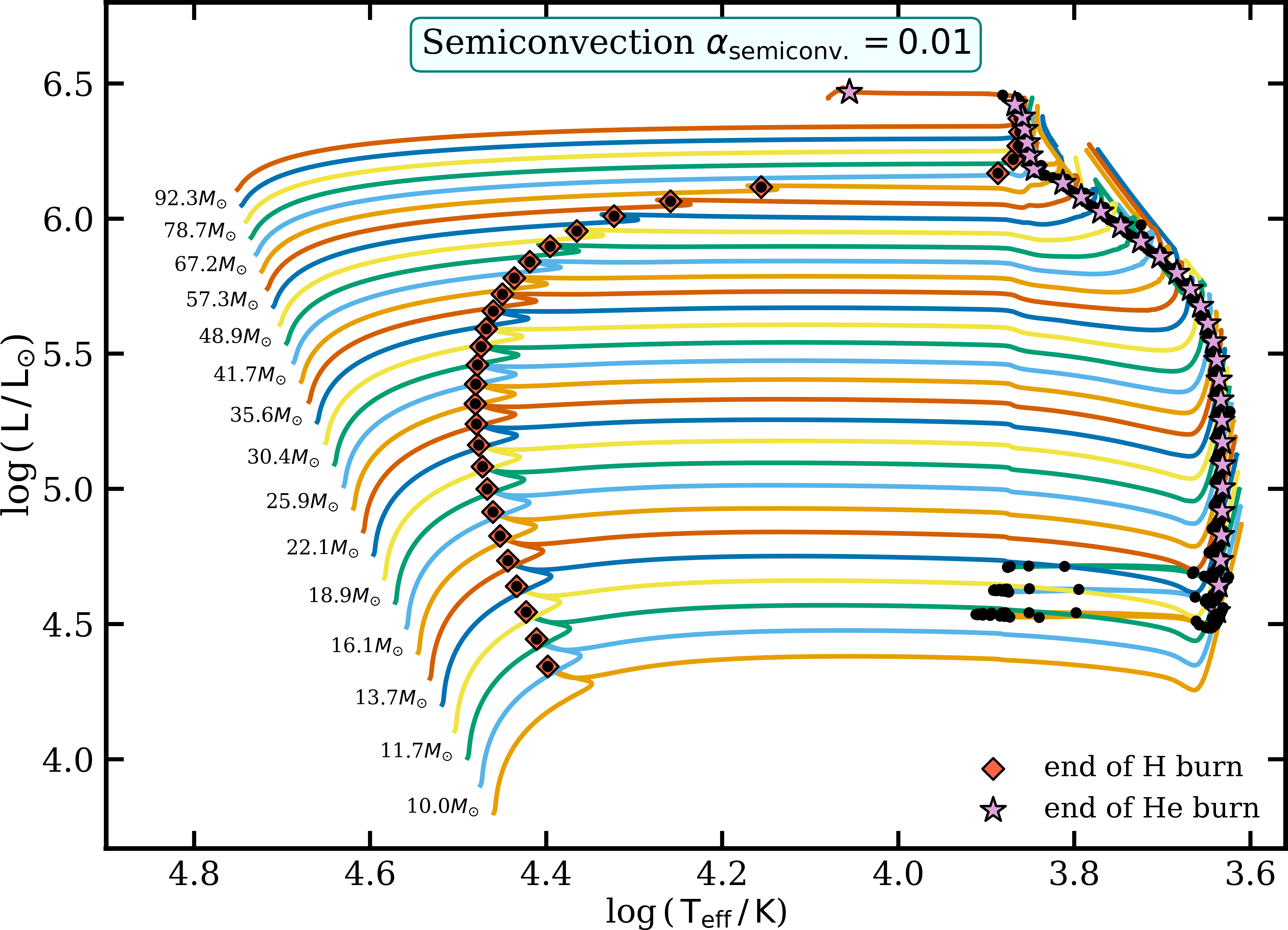}
    \caption{Hertzsprung Russell diagram with single stellar tracks that were used as a basis for the computation of critical mass ratios, minimum orbital separations from SMT evolution, and the corresponding minimum BBH merger delay times (see results for stellar models with semiconvection $\alpha_{\rm semiconv.} = 33$ in Fig.~\ref{fig.triumvirate} and with $\alpha_{\rm semiconv.} = 0.01$ in the right panel of Fig.~\ref{fig.obywatel_gc}). Here there tracks are of 30 different masses, logarithmically spaced from $10$ to $100\Msun$, evolved until at least the end of core-He burning. Black dots are equally spaced every 50,000 years during the post-MS evolution.}
    \label{fig.HRD}
\end{figure*}

\begin{figure*}
    \includegraphics[width=\textwidth]{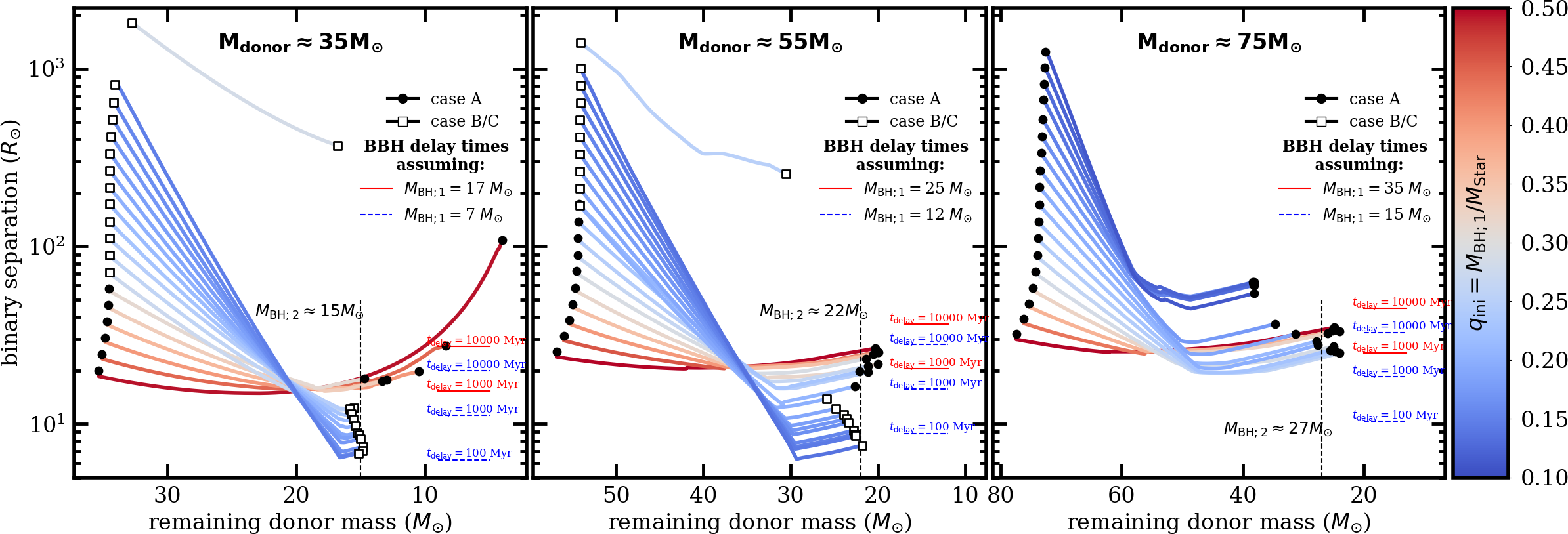}
    \caption{Evolution of binary separation during the mass transfer phase in MESA models of BH+O-star systems with $q = q_{\rm crit}$ mass ratios and initial O-star masses of $35$, $55$, and $75\Msun$ (the three different panels), plotted as a function of the decreasing remaining donor mass.
    Different lines correspond to different initial separations, colored by the initial mass ratio $M_{\rm donor} / M_{\rm accretor}$. 
    The initially more narrow systems interact when the donor is a MS star (closed circles), whereas in wider orbits the interaction is from a post-MS donor (open squares). Because the mass ratio is critical for each mass and radius of the donor at RLOF, the models shown here lead to the smallest binary separations via SMT evolution given the donor. See Fig.~\ref{fig.triumvirate} for a full parameter space exploration, with other initial donor masses and radii.
    Assuming that at the end of its evolution the donor collapses to form a BH with mass $15$, $22$, and $27\Msun$ (for different panels), we mark which binary separations would be required for BBH delay times of $100$, $1000$, or $10000$Myr for BBH mass ratio $0.9$. 
    Counter-intuitively, the smallest separations and the shortest BBH delay times from the SMT channel may originate from initially the widest BH+O-star systems (post-MS donors in the $35$
    and $55 \Msun$ panels).
}
    \label{fig.slice}
\end{figure*}

\begin{figure*}
    \includegraphics[width=\textwidth]{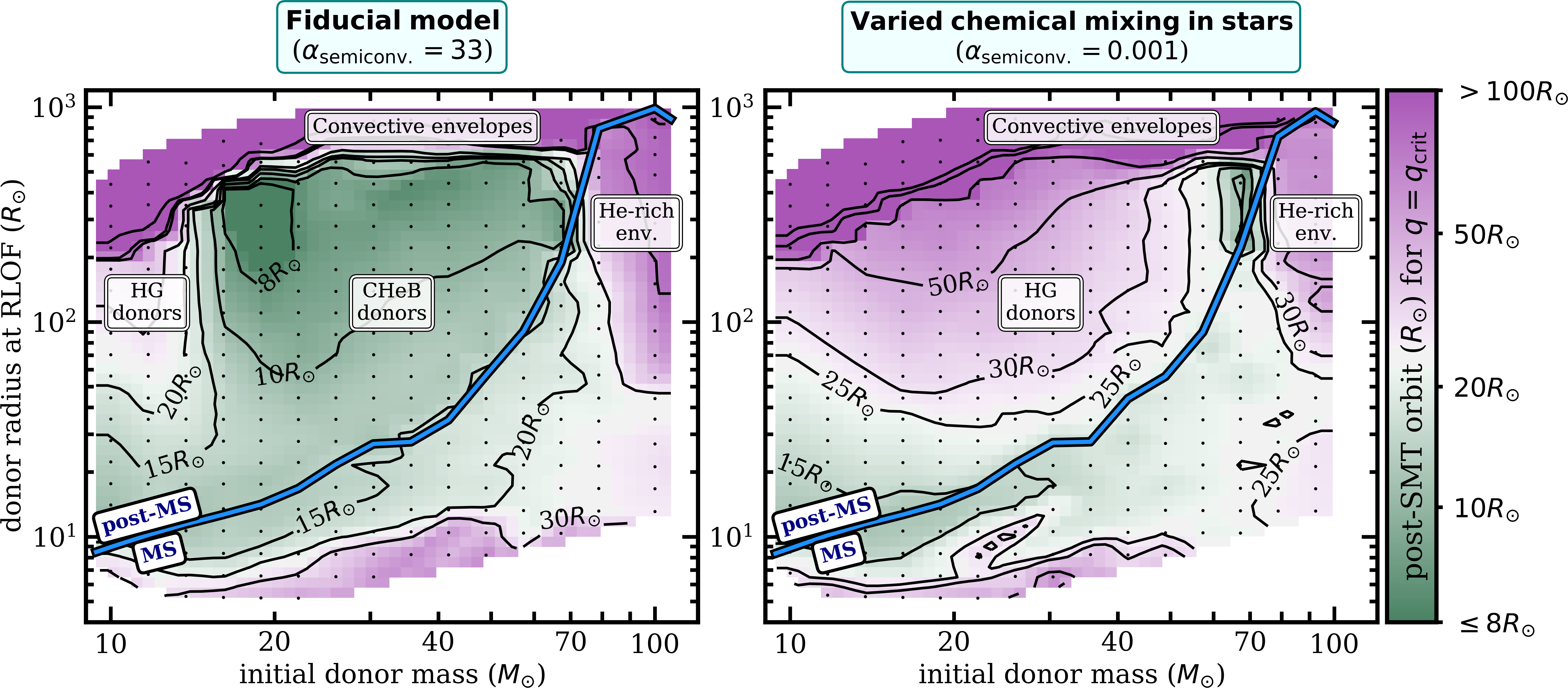}
    \caption{Same as Fig.~\ref{fig.obywatel_gc} but comparing the final orbital separations instead of the BBH delay times. In the variation with low efficiency of semiconvection (right) all the post-MS donors are of the Hertzsprung Gap type (HG). They are characterized by wider post-SMT separations compared to core-He burning (CHeB) donors due to differences in their internal chemical / entropy structure (Fig.~\ref{fig.disc_different_donors}). Based on that, we propose a new mass transfer stability treatment in rapid binary codes in Sec.~\ref{sec.disc_stability_treatment}. The difference between stellar models in both panels is closely connected to the so-called blue-supergiant problem (Sec.~\ref{sec.disc_GW_stellar_interiors}). } 
    \label{fig.app_minSep_lowSC}
\end{figure*}

 \end{appendix}

\end{document}